\documentclass[preprint,sort&compress,11pt,3p]{elsarticle}

\usepackage{moreverb}
\usepackage{mathtools}
\usepackage{amsmath}
\usepackage{amssymb}
\usepackage{algorithmic}
\usepackage{graphics}
\usepackage{graphicx}
\usepackage{subfigure}
\usepackage{caption}
\usepackage{overpic}
\usepackage{enumitem}
\usepackage{extarrows}
\usepackage{color}
\usepackage{framed}
\usepackage{wrapfig}
\usepackage{bm}
\usepackage{mathrsfs}
\usepackage{mathabx}
\usepackage{multirow}
\usepackage{longtable}
\usepackage{hyperref}
\usepackage{paralist}
\usepackage{indentfirst}
\usepackage{relsize}
\usepackage{extarrows}
\usepackage{lineno,xcolor}
\usepackage{upgreek}
\usepackage{bm}
\usepackage{mwe}
\usepackage{xcolor}
\usepackage{booktabs}
\usepackage[flushleft]{threeparttable}
\usepackage[linesnumbered,lined,ruled]{algorithm2e}

\newcommand{\eref}[1]{(\ref{#1})}

\hypersetup{
bookmarks=true,
bookmarksopen=true,
bookmarksnumbered=true,
unicode=false,
pdftoolbar=true,
pdfmenubar=true,
pdffitwindow=false,
pdfstartview={FitH},
pdftitle={My title},
pdfauthor={Author},
pdfsubject={Subject},
pdfcreator={Creator},
pdfproducer={Producer},
pdfkeywords={keywords},
pdfnewwindow=true,
colorlinks=true,
linkcolor=blue,
citecolor=blue,
filecolor=magenta,
urlcolor=blue
}

\journal{Elsevier}

\DeclareMathOperator*{\argmin}{arg\,min}

\begin{document}
	
\title{\Large SeismicNet: Physics-informed neural networks for seismic wave modeling in semi-infinite domain}

\author[NU1]{Pu Ren\fnref{equal}}
\ead{ren.pu@northeastern.edu}
\author[NU2]{Chengping Rao\fnref{equal}}
\author[BJUT]{Su Chen}
\author[ND]{Jian-Xun Wang}
\author[RUC1,RUC2]{Hao Sun\corref{cor}}
\ead{haosun@ruc.edu.cn}
\author[UCAS]{Yang Liu\corref{cor}}
\ead{liuyang22@ucas.ac.cn}

\cortext[cor]{Corresponding authors.}
\fntext[equal]{Equal contribution.}

\address[NU1]{Department of Civil and Environmental Engineering, Northeastern University, Boston, MA 02115, USA}
\address[NU2]{Department of Mechanical and Industrial Engineering, Northeastern University, Boston, MA 02115, USA}
\address[BJUT]{Key Laboratory of Urban Security and Disaster Engineering of the Ministry of Education, Beijing University of Technology, Beijing, 100124, China}
\address[ND]{Department of Aerospace and Mechanical Engineering, University of Notre Dame, Notre Dame, IN 46556, USA}
\address[RUC1]{Gaoling School of Artificial Intelligence, Renmin University of China, Beijing, 100872, China}
\address[RUC2]{Beijing Key Laboratory of Big Data Management and Analysis Methods, Beijing, 100872, China}
\address[UCAS]{School of Engineering Sciences, University of Chinese Academy of Sciences, Beijing, 101408, China}

\begin{abstract}
\small
Recently, there has been an increasing interest in leveraging physics-informed neural networks (PINNs) for modeling dynamical systems. However, very limited studies have been conducted along this horizon on seismic wave modeling tasks. A critical challenge is that these geophysical problems are typically defined in large domains (i.e., semi-infinite), which leads to high computational cost. In this paper, we present a new PINN model for seismic wave modeling in semi-infinite domain without the need of labeled data. In specific, the absorbing boundary condition is introduced into the network as a soft regularizer for handling truncated boundaries. To scale up, we consider a sequential training strategy via temporal domain decomposition to improve the scalability of the network and solution accuracy. Moreover, we design a novel surrogate modeling strategy to account for parametric loading, which estimates the wave propagation in semi-infinite domain given the seismic loading at different locations. Various numerical experiments have been implemented to evaluate the performance of the proposed PINN model in the context of forward modeling of seismic wave propagation. In particular, we define diverse material distributions to test the versatility of this approach. The results demonstrate excellent solution accuracy under distinctive scenarios.
\end{abstract}


\maketitle

\section{Introduction}\label{s:intro}
Understanding seismic wave propagation is crucial for the communities of seismology and earthquake engineering. Typically, we record the seismic signals with seismometers and derive various information about the composition of the Earth, e.g., the discovery of Mohorovicic discontinuity which separates the crust and the mantle of the Earth. For decades, we have observed a significant advancement in the theoretical investigation of seismic wave physics. Furthermore, the theoretical breakthrough also brings about the rapid development of numerical modeling of seismic wave propagation for applied seismology and engineering practices, such as finite difference~\cite{kelly1976synthetic,virieux1986p}, finite element~\cite{bao1998large} and spectral element~\cite{komatitsch1998spectral,komatitsch1999introduction} methods. In general, the traditional numerical methods employ finite sets of basis functions (e.g., polynomials) and the corresponding parameters to approximate the analytical solutions of differential equations. They excel in solution accuracy for various ordinary or partial differential equations (ODEs/PDEs), but may show weakness in surrogate modeling and inverse analysis, which require repeated runs of forward models and lead to high computational costs.

Thanks to the great progress of artificial intelligence (AI), in particular machine learning (ML), many scientists start to explore the potential of using ML to address seismic wave propagation problems~\cite{ross2018p,zhu2019phasenet,bergen2019machine,kong2019machine,mousavi2020earthquake,mousavi2022deep}. For example, Fourier Neural Operator (FNO)~\cite{li2020fourier} has been investigated for forward and inverse analysis of 2D acoustic wave equation~\cite{yang2021seismic}. Nevertheless, these research efforts rely on large amounts of high-fidelity labeled data, which are usually difficult to record and collect. It has been a pressing task to develop novel ML methods for dealing with imperfect measurement data (e.g., sparse and potentially noisy).

The recent development of PINNs~\cite{raissi2019physics,karniadakis2021physics} has shed new light on scientific computing with limited labeled data. The general principle of PINNs is to integrate deep neural networks (DNNs) and physical laws to learn the underlying consistent dynamics from small or zero labeled data. The advent of PINNs provides a new perspective for training neural networks, where the introduction of physical laws into the network strengthens the constraint for optimization in small data regimes. For instance, the latest research works~\cite{zhu2019physics,sun2020surrogate,geneva2020modeling,gao2021phygeonet,rao2021physics,ren2022phycrnet} attempt to solve PDE systems when the full physics knowledge is accessible, i.e., governing equations, initial and boundary conditions (I/BCs). In addition, we observe the explosive research of applying this paradigm to miscellaneous scientific domains, including fluid flows~\cite{wang2020towards,raissi2020hidden,rao2020physics,erichson2019physics,jin2021nsfnets,wang2022respecting,tartakovsky2020physics}, solid mechanics~\cite{zhang2020physics,zhang2020physicslstm,haghighat2021physics,chen2021learning}, multi-scale and multi-physics modeling~\cite{cai2021deepm,lin2021operator,mao2021deepm}, heat transfer~\cite{cai2021physics}, equation discovery~\cite{raissi2018deep,sun2021physics,chen2021physics,rao2022discovering} and data augmentation~\cite{esmaeilzadeh2020meshfreeflownet,rao2021embedding,ren2022physics,gao2021super}. Although PINN has shown great success in various scientific problems, it is also noteworthy that they are not competent in comparison with traditional numerical methods for forward simulations in terms of solution accuracy. Their huge potential lies in tackling surrogate modeling and inverse analysis with a tradeoff between accuracy and computational efficiency, especially when only sparse data are recorded.

In this paper, we present a PINN approach for forward modeling of seismic wave propagation, as a basis for possible inverse analysis (e.g., full waveform inversion, a.k.a., FWI). Recently, a few studies~\cite{bin2020eikonal,smith2020eikonet,song2021solving,rasht2022physics} show the efficacy of PINN succeeding in solving FWI problems in the context of acoustic wave equations. Note that the acoustic wave equations are formulated in scalar field and generally used to model wave propagation under deep subsurface (e.g., at the km scale), which are inapplicable to near-surface waveform simulation and inversion. Elastic wave equations provide a solution to near sub-surface wave propagation modeling. To this end, we establish a PINN model for solving elastic wave equations (i.e., a more complex PDE system) in semi-infinite domain with inhomogeneous material profiles. We also design a PINN-based surrogate model to directly infer the full waveforms given diverse loading locations, which, to the best of our knowledge, is the first attempt in the context of seismic wave propagation modeling. The main contributions of this paper are three-fold. First, we introduce the absorbing boundary condition (ABC) into PINN to deal with truncated boundaries in semi-infinite domain, which works as a soft regularizer in network optimization, to enable reliable wave progragation. Second, a temporal decomposition training strategy is presented to improve the scalability and solution accuracy of our PINN architecture for large-scale modeling. Lastly, we validate the effectiveness of our approach for elastic wave propagation modeling compared with the reference numerical results under various scenarios (e.g., different material distributions and loading conditions).

The rest of the paper is organized as follows, apart from this Introduction section. Section \ref{s:prob} formulates the scientific problem of forward analysis of seismic wave propagation. Section \ref{s:method} presents the details of the specific PINN model, an introduction to ABC, the sequential training scheme via domain decomposition, and the strategy for parametric surrogate modeling. In Section \ref{s:experiments}, we show the results for extensive numerical experiments to evaluate the performance of our proposed method. Section \ref{s:discussion} discusses the advantages and limitations of the current method as well as the future research directions. Section \ref{s:conclusion} concludes the entire paper.

\section{Problem formulation}\label{s:prob}
In this paper, we aim to investigate the potential of PINN for seismic wave propagation modeling in 2D elastic media. The specific governing equations of interest are given by 
\begin{equation} 
\label{eq:seismic_wave_pde} 
\begin{aligned}
    \rho\frac{\partial^2 u_0}{\partial t^2}&=\frac{\partial}{\partial x}\left ( (2\mu+\lambda)\frac{\partial u_0}{\partial x} + \lambda\frac{\partial u_1}{\partial y} \right ) + \frac{\partial}{\partial y} \left ( \mu(\frac{\partial u_1}{\partial x} + \frac{\partial u_0}{\partial y}) \right ),\\
    \rho\frac{\partial^2 u_1}{\partial t^2}&= \frac{\partial}{\partial x} \left ( \mu(\frac{\partial u_1}{\partial x} + \frac{\partial u_0}{\partial y}) \right ) + \frac{\partial}{\partial y}\left ( \lambda\frac{\partial u_0}{\partial x} + (2\mu+\lambda)\frac{\partial u_1}{\partial y} \right ), 
\end{aligned}
\end{equation}
%
where $\{x,y\}$ and $t$ denote the spatial and temporal coordinates, respectively; $\{u_0,u_1\}$ are the two displacement variables (horizontal and vertical). $\rho$ is the density and $\{\lambda,\mu\}$ represent Lam\'e constants of the media. Regarding the setting of boundary conditions (BCs), we provide a detailed illustration in Section~\ref{s:abc}. In addition, Eq.~\eref{eq:seismic_wave_pde} can also be rewritten in a matrix formulation, which is summarized as:
\begin{equation} 
\label{eq:seismic_wave_pde2} 
    \mathbf{{u}}_{tt} = \mathbf{D}_1 \mathbf{{u}}_{xx} + \mathbf{H} \mathbf{{u}}_{xy} + \mathbf{D}_2 \mathbf{{u}}_{yy}.
\end{equation}
where $\mathbf{u}=[u_0,u_1]^{\top}$ is the displacement vector and the subscripts denote the partial derivatives, e.g., $\mathbf{u}_{xx}=\partial^2 \mathbf{u} / {\partial x^2}$. Moreover, $\{\mathbf{D}_1,\mathbf{D}_2,\mathbf{H}\}$ refer to the coefficient matrices associated with material properties, which are expressed as 
\begin{equation}
\label{eq:coeff_matrix}
\mathbf{D}_1=\begin{bmatrix}
\alpha^2 & 0  \\
0 & \beta^2 
\end{bmatrix}, \;\;
\mathbf{D}_2=\begin{bmatrix}
\beta^2  & 0  \\
0 & \alpha^2
\end{bmatrix}, \;\;
\mathbf{H}=\begin{bmatrix}
0 & \alpha^2-\beta^2 \\
\alpha^2-\beta^2 & 0
\end{bmatrix}.
\end{equation}
Herein, $\alpha$ and $\beta$ represent the longitudinal and transverse wave speeds, respectively, which can be calculated through
\begin{equation} 
\label{eq:lame2wavespeed} 
    \alpha=\sqrt{\frac{\lambda+2\mu}{\rho}}, \;\;
    \beta=\sqrt{\frac{\mu}{\rho}}.
\end{equation}

Our objective is to leverage PINN for forward simulation of elastic wave propagation, including solving the specific PDEs and parametric surrogate modeling. To be more precise, we aim to solve the seismic wave propagation equations in elastic media (as shown in Eq.~\eref{eq:seismic_wave_pde} or \eref{eq:seismic_wave_pde2}) using PINN in a given semin-infinite domain. We only rely on the accessible physical principles (i.e., governing equations and I/BCs) to optimize the entire network without any labeled data. In the part of parametric surrogate modeling, we build a PINN architecture to extrapolate the seismic responses with respect to different loading positions. Herein, a new variable $\mathbf{x}_c$ is introduced into PINNs apart from the typical coordination information (e.g., $\{t,x,y\}$). Here, $\mathbf{x}_c$ is used to mark the specific spatial loading location of interest, which serves for both training and extrapolation.

\section{Methodology}\label{s:method}
In this section, we introduce a new PINN architecture for forward modeling of elastic wave propagation in semin-infinite domain. Considering the characteristics of seismological tasks, the ABC is applied on the truncated boundaries to eliminate wave reflection due to boundary effect. A domain decomposition training strategy is then introduced to handle large-scale simulations. Furthermore, we also propose a specifically-designed DNN-based framework for parametric surrogate modeling, which is slightly different from vanilla PINN used for solving generic PDEs.

\subsection{PINNs}\label{s:dnn}

DNNs have received considerable attention in the scientific computing community thanks to their theoretical foundation of universal approximation~\cite{cybenko1989approximation,hornik1989multilayer}. The previous implementations of DNNs for scientific tasks generally rely on a large amount of labeled data. Nevertheless, the recent breakthrough of PINNs~\cite{raissi2019physics} has enabled learning in a ``small data'' regime by incorporating physical constraints (e.g., governing equations and I/BCs) into the networks. Typically, a PINN architecture is composed of an input layer, multiple hidden layers and an output layer. Each layer is designated with a specified neuron number. For instance, the mathematical formulation of the connection between the $(i-1)$-th layer and the $i$-th layer is given by  
\begin{equation}
    \label{eq:dnn}
    \mathbf{X}^i = \sigma \left(\mathbf{W}^i \mathbf{X}^{i-1} + \mathbf{b}^i \right),\;\; i \in [1,n+1],
\end{equation}
where $\{\mathbf{W}^i,\mathbf{b}^i\}$ denote the corresponding weight matrix and bias vector at the $i$-th layer, respectively; $\sigma(\cdot)$ is a nonlinear activation function; $n$ is the number of hidden layers. Here, $\{\mathbf{X}^{i-1},\mathbf{X}^i\}$ are the input and output variables at the $i$-th layer. The general goal of PINNs is to leverage DNNs to approximate the solutions of PDEs of interest with a spatial or spatiotemporal input coordinates (e.g., $\{t,x,y\}$ for 2D time-dependent dynamics). Herein, the difference between general DNN-based solvers and PINNs lies in the introduction of physics loss apart from data loss. The physics loss is obtained based on the strong form of PDEs and the derivative terms of interest are calculated via automatic differentiation~\cite{baydin2018automatic}. We discuss the specific network architecture for our tasks in Section~\ref{s:network}.

\begin{figure}[b!]
	\centering
	\includegraphics[width=0.45\textwidth]{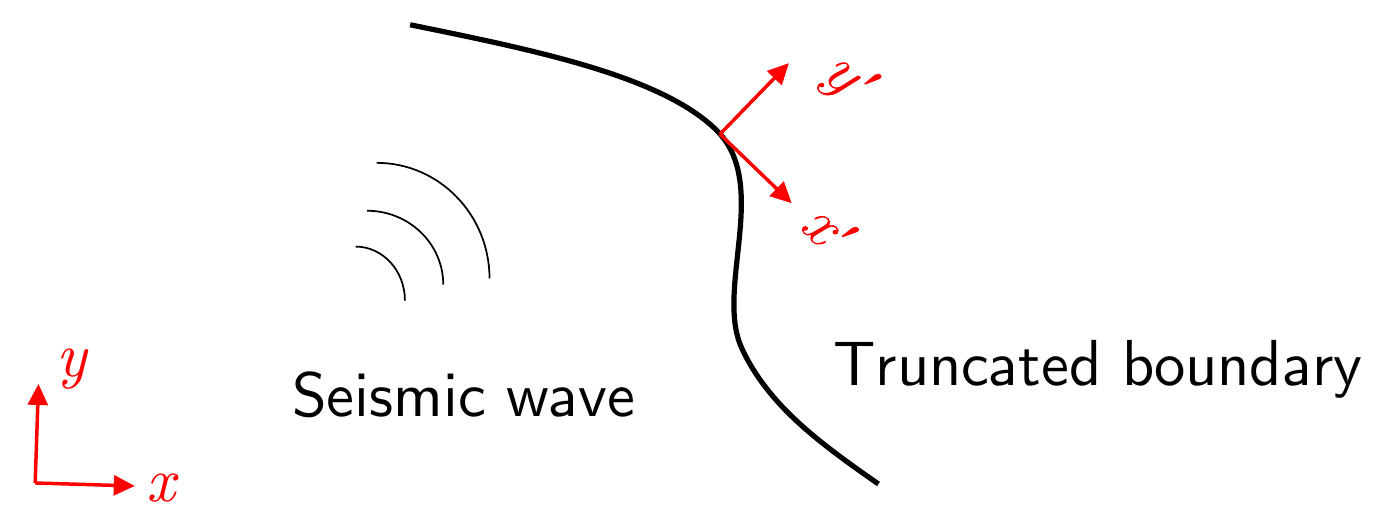}
	\caption{Illustrative diagram of the absorbing boundary condition. The local coordinate system is determined by $x'$ (tangential direction) and $y'$ (outer normal direction) axes. $x$ and $y$ axes define the global coordinate.}
	\label{fig:seismic_wave:abc}
\end{figure}

\subsection{Absorbing boundary condition}\label{s:abc}
As the seismic wave propagation problem commonly features a semi-infinite domain (e.g., in Earth), the computational domain needs to be truncated to avoid unnecessary computation. Nevertheless, one of the main obstacles is to guarantee that the wave propagates out the computational domain without interfering with the upstream field when it passes through truncated boundaries. To overcome this issue in numerical simulations, researchers have developed various mathematical formulations \cite{engquist1977absorbing,clayton1977absorbing,engquist1979radiation,givoli1990non} to describe ABCs on the truncated boundaries. The illustrative diagram of the ABC is presented in Fig.~\ref{fig:seismic_wave:abc}. In this paper, we consider one of the most popular ABCs, which is given by
\begin{equation} 
\label{eq:seismi:abc} 
    \mathbf{u}_{ty'} + \mathbf{C}_1 \mathbf{u}_{tt} + \mathbf{C}_2 \mathbf{u}_{tx'} + \mathbf{C}_3 \mathbf{u}_{x'x'} = \mathbf{0}.
\end{equation}
where $x'$ (tangential direction) and $y'$ (outer normal direction) axes determine the local coordinate system. In addition, $\{\mathbf{C}_1,\mathbf{C}_2,\mathbf{C}_3\}$ are coefficients matrices reflecting the material properties of the media, written as 
\begin{equation}
\label{eq:C1C2C3}
    \mathbf{C}_1=\begin{bmatrix}
    1/\beta & 0  \\
    0 & 1/\alpha
    \end{bmatrix}, \;\;
    \mathbf{C}_2=(\beta-\alpha)\begin{bmatrix}
    0 & 1/\beta   \\
    1/\alpha & 0 
    \end{bmatrix}, \;\;
    \mathbf{C}_3=\frac{1}{2}\begin{bmatrix}
    \beta-2\alpha & 0   \\
    0 & \alpha - 2\beta
    \end{bmatrix}.
\end{equation}

Note that the ABC described in Eq.~\eref{eq:seismi:abc} is defined by using the local coordinate system. Therefore, a proper coordinate transformation is required for automatic differentiation in PINNs. By introducing the ABC into the network, we are free from simulating a large computational domain with PINNs. Moreover, since the PINN deals with the strong form of the governing equation, Eq.~\eref{eq:seismi:abc} can be used readily to impose the corresponding BCs. In contract, the variational form of the analytical law is required in traditional finite element analysis, which brings additional complexity in numerical implementation.

\subsection{Network architecture}\label{s:network}
First of all, we observe that Eq.~\eref{eq:seismic_wave_pde} is characterized by many second-order derivatives, which could cause a huge computational graph for PINN. Our previous work \cite{rao2020physics} has demonstrated that reducing the order of automatic differentiation (i.e., the order of derivatives in governing equations) can effectively improve the learnability and the solution accuracy of PINNs. Hence, we re-formulate the governing equations in Eq.~\eref{eq:seismic_wave_pde} to an equivalent state-space form, which are given by
\begin{equation} 
\label{eq:seismic_elasto} 
\begin{aligned}
    \rho \mathbf{v}_t &= \mathbf{Q}\bm{\sigma}, \\
    \bm{\sigma}_t &= \mathbf{C} \bm{\epsilon}_t, \\
    \mathbf{u}_t &= \mathbf{v}.
\end{aligned}    
\end{equation}
where $\mathbf{u}=[u_0,u_1]^\top$ and $\mathbf{v}=[v_0,v_1]^\top$ refer to the displacement and velocity vectors, respectively. $\{\bm{\sigma},\bm{\epsilon}_t\}$ denote the corresponding stress and strain rate vectors, which are expressed as
\begin{equation}
\label{eq:sigma_epsilon}
\begin{aligned}
    \bm{\sigma} &= \left[\sigma_{xx}, \;\sigma_{yy}, \; \sigma_{xy}\right]^{\top}, \\
    \bm{\epsilon}_t &= \left[\frac{\partial v_0}{\partial x}, \; \frac{\partial v_1}{\partial y}, \; \frac{\partial v_0}{\partial y} + \frac{\partial v_1}{\partial x}\right]^\top.
\end{aligned}    
\end{equation}
Note that $\{\sigma_{xx},\sigma_{yy}\}$ represent the normal stresses along $x$-dimension and $y$-dimension, respectively; $\sigma_{xy}$ is the shear stress. In addition, $\mathbf{Q}$ is an operator matrix and $\mathbf{C}$ represents the constitutive tensor, which are formulated as 
\begin{equation}
\label{eq:Q_C}
\begin{aligned}
    \mathbf{Q} &= {\begin{bmatrix}
    \frac{\partial}{\partial x} & 0 & \frac{\partial}{\partial y} \\[0.4em]
    0 & \frac{\partial}{\partial y} & \frac{\partial}{\partial x} 
    \end{bmatrix}}, \\
    \mathbf{C} &= {\begin{bmatrix}
    \lambda +2\mu & \lambda & 0 \\
    \lambda & \lambda +2\mu & 0 \\
    0 & 0 & \mu \\
    \end{bmatrix}}.
\end{aligned}    
\end{equation}
In essence, Eq.~\eref{eq:seismic_elasto} represent the governing equations of elastodynamics under the assumption of plane strain \cite{reddy2017energy}.

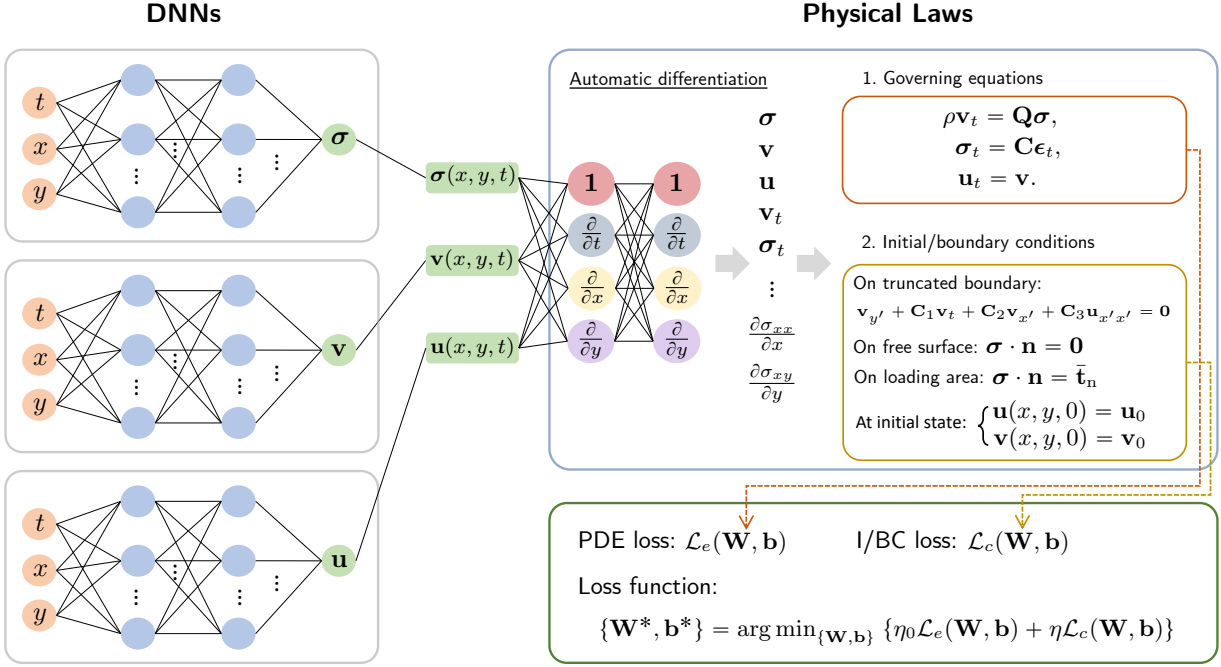
\begin{figure}[t!]
	\centering
	\begin{overpic}[width=0.99\textwidth]{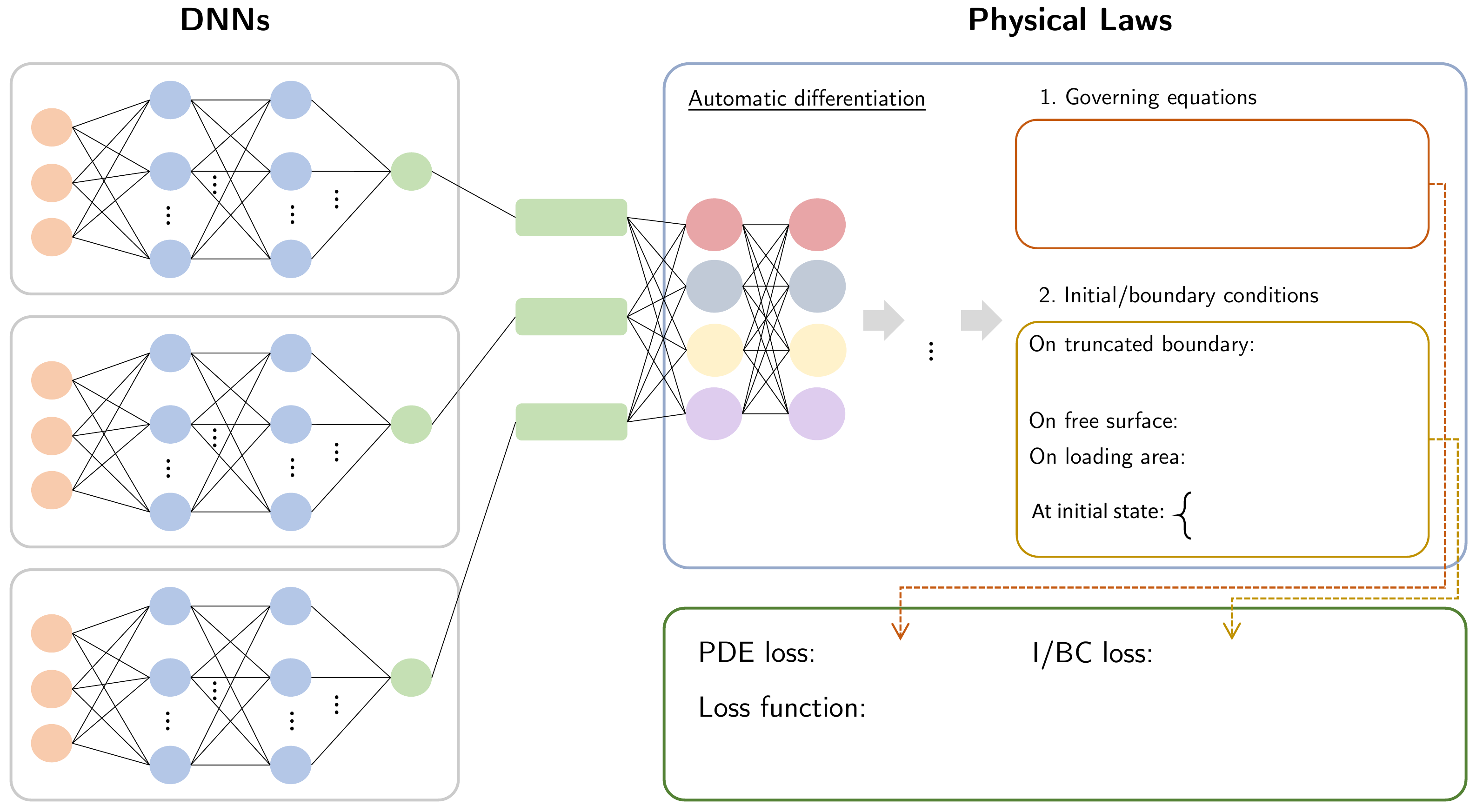}
	\put(3.2,45.8){\color{black}{\small $t$}}
	\put(3.0,42.2){\color{black}{\small $x$}}
	\put(3.0,38.6){\color{black}{\small $y$}}
	
	\put(3.2,28.6){\color{black}{\small $t$}}
	\put(3.0,25.0){\color{black}{\small $x$}}
	\put(3.0,21.4){\color{black}{\small $y$}}	

	\put(3.2,11.5){\color{black}{\small $t$}}
	\put(3.0,7.9){\color{black}{\small $x$}}
	\put(3.0,4.3){\color{black}{\small $y$}}
	
	\put(27.2,43){\color{black}{\small $\boldsymbol{\sigma}$}}
 	\put(27.2,25.8){\color{black}{\small $\mathbf{v}$}}
 	\put(27.2,8.6){\color{black}{\small $\mathbf{u}$}}

	\put(35.3,40){\color{black}{\scriptsize $\boldsymbol{\sigma}(x,y,t)$}}
 	\put(35.3,33.2){\color{black}{\scriptsize $\mathbf{v}(x,y,t)$}}
 	\put(35.3,26){\color{black}{\scriptsize $\mathbf{u}(x,y,t)$}}

	\put(47.8,39.2){\color{black}{\small $\mathbf{1}$}}
 	\put(47.4,35.2){\color{black}{\small $\frac{\partial}{\partial t}$}}
 	\put(47.3,30.8){\color{black}{\small $\frac{\partial}{\partial x}$}}
 	\put(47.2,26.6){\color{black}{\small $\frac{\partial}{\partial y}$}}

	\put(54.7,39.2){\color{black}{\small $\mathbf{1}$}}
 	\put(54.3,35.2){\color{black}{\small $\frac{\partial}{\partial t}$}}
 	\put(54.2,30.8){\color{black}{\small $\frac{\partial}{\partial x}$}}
 	\put(54.2,26.6){\color{black}{\small $\frac{\partial}{\partial y}$}}
	
	\put(62,44.6){\color{black}{\small $\boldsymbol{\sigma}$}}
	\put(62,42){\color{black}{\small $\mathbf{v}$}}
	\put(62,39.4){\color{black}{\small $\mathbf{u}$}}
	\put(62,36.8){\color{black}{\small $\mathbf{v}_t$}}
	\put(62,34.2){\color{black}{\small $\boldsymbol{\sigma}_t$}}	
	\put(61,27){\color{black}{\small $\frac{\partial \sigma_{xx}}{\partial x}$}}	
	\put(61,23){\color{black}{\small $\frac{\partial \sigma_{xy}}{\partial y}$}}		
	\put(77,44.7){\color{black}{\footnotesize $\rho \mathbf{v}_t = \mathbf{Q}\bm{\sigma},$}}
	\put(78,42.2){\color{black}{\footnotesize $\bm{\sigma}_t = \mathbf{C} \bm{\epsilon}_t,$}}	
	\put(78.2,39.7){\color{black}{\footnotesize $\mathbf{u}_t = \mathbf{v}.$}}		
	\put(70,29){\color{black}{\tiny $\mathbf{v}_{y'} + \mathbf{C}_1 \mathbf{v}_{t} + \mathbf{C}_2 \mathbf{v}_{x'} + \mathbf{C}_3 \mathbf{u}_{x'x'} = \mathbf{0}$}}
	
	\put(80.5,26){\color{black}{\footnotesize $\bm{\sigma} \cdot \mathbf{n} = \mathbf{0}$}}
	\put(81,23.5){\color{black}{\footnotesize $\bm{\sigma} \cdot \mathbf{n} = \bar{\mathbf{t}}_\text{n}$}}
	\put(81,20.7){\color{black}{\footnotesize $\mathbf{u}(x,y,0)=\mathbf{u}_0$}}
	\put(81.1,18.5){\color{black}{\footnotesize $\mathbf{v}(x,y,0)=\mathbf{v}_0$}}
	
	\put(56,10.1){\color{black}{\footnotesize $\mathcal{L}_e(\mathbf{W},\mathbf{b})$}}	
	\put(79,10.1){\color{black}{\footnotesize $\mathcal{L}_c(\mathbf{W},\mathbf{b})$}}
	\put(49,3){\color{black}{\footnotesize $\{\mathbf{W}^*,\mathbf{b}^*\}=\argmin_{\{\mathbf{W},\mathbf{b}\}} \; \{\eta_0\mathcal{L}_{e}(\mathbf{W},\mathbf{b})+\eta \mathcal{L}_{c}(\mathbf{W},\mathbf{b})\}$}}

	\end{overpic}
	\caption{Overview of the proposed PINN framework for seismic wave propagation, including the network architecture and the constraint of physical laws. Three separate DNNs are used to approximate the displacement $\mathbf{u}$, velocity $\mathbf{v}$, and stress $\bm{\sigma}$, respectively. Automatic differentiation is exploited for obtaining the derivative terms and further constructing the loss function.} 
	\label{fig:seismic_wave:PINN_architecture}
\end{figure}

Furthermore, the overall architecture of our designed PINNs for modeling seismic wave propagation is presented in Fig.~\ref{fig:seismic_wave:PINN_architecture}. Three separate DNNs are employed to approximate the variables of displacement $\mathbf{u}$, velocity $\mathbf{v}$, and stress $\bm{\sigma}$ considering their different numerical scales. After obtaining these field variables, we utilize the automatic differentiation \cite{baydin2018automatic} to evaluate the partial derivatives of interests with respect to the input coordinates. Based on the differential terms, we construct a loss function $\mathcal{L}(\mathbf{W},\mathbf{b})$ of the concerned physical laws in PINN, including governing equations and I/BCs (including the ABCs). In specific, the equation loss $\mathcal{L}_e(\mathbf{W},\mathbf{b})$ is a mean square error (MSE) loss that follows the analytical expressions of elastodynamics as shown in Eq.~\eref{eq:seismic_elasto}. Such a loss component helps preserve the inherent PDE structure. The I/BC loss $\mathcal{L}_c(\mathbf{W},\mathbf{b})$ is also an MSE loss that measures the misfit in the context of given I/BCs. It is noteworthy that this type of PINNs does not rely on observation data loss in forward simulation. Namely, only equation loss $\mathcal{L}_e$ and I/BC loss $\mathcal{L}_c$ are sufficient to constrain the network and also respect the consistent physical principles. The optimal network parameters $\{\mathbf{W}^*,\mathbf{b}^*\}$ are learned by minimizing the combined loss function, as follows
\begin{equation}
\label{eq:loss_func}
\{\mathbf{W}^*,\mathbf{b}^*\}=\argmin_{\{\mathbf{W},\mathbf{b}\}} \; \{\eta_0\mathcal{L}_{e}(\mathbf{W},\mathbf{b})+\eta \mathcal{L}_{c}(\mathbf{W},\mathbf{b})\},
\end{equation}
where $\eta_0$ and $\eta$ are weighting coefficients (i.e., user-defined hyperparameters).

\subsection{Sequential training via domain decomposition}\label{s:domain_decomp}
\begin{figure}[t!]
	\centering
	\includegraphics[width=0.4\textwidth]{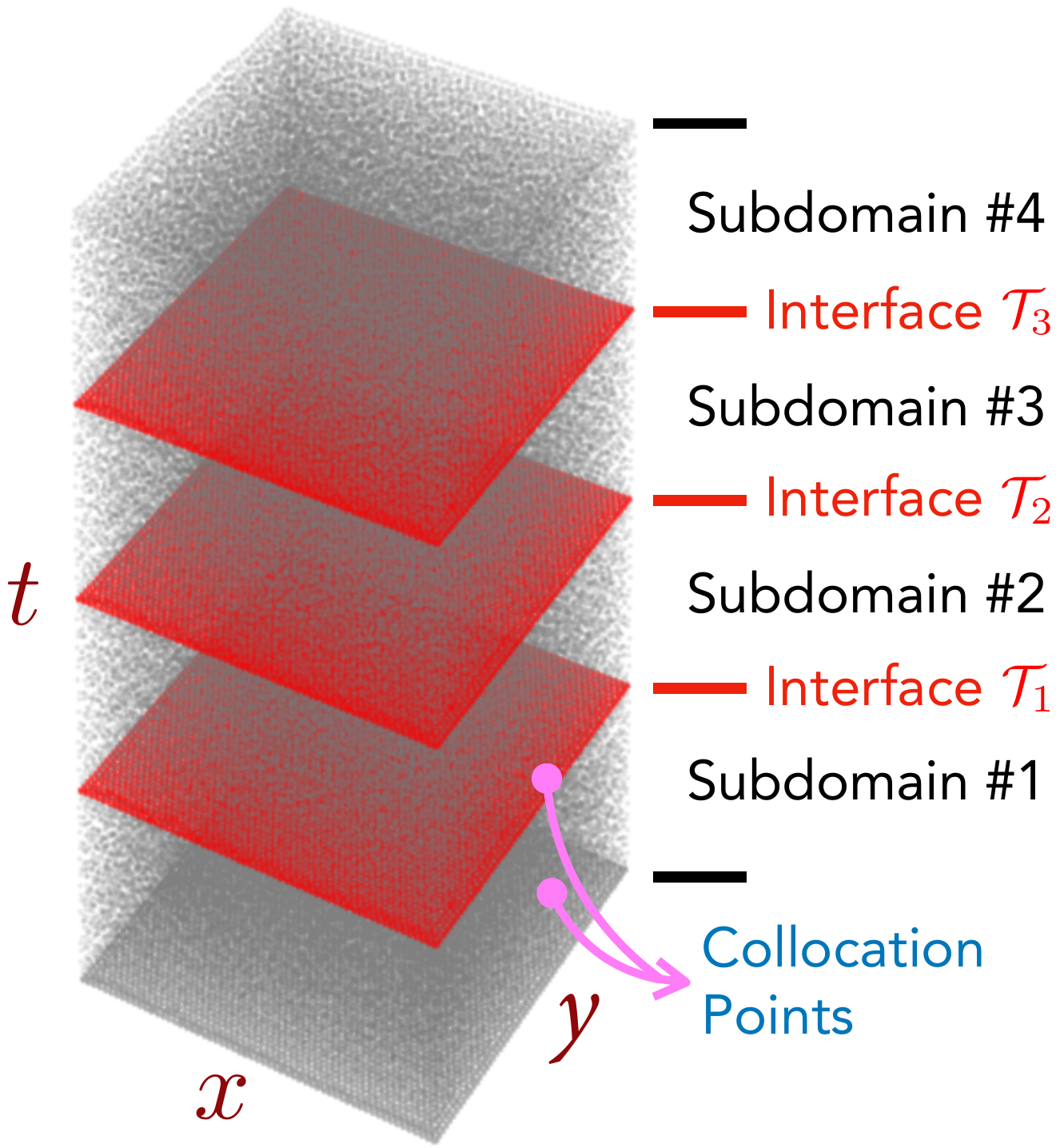}
	\caption{Schematic of temporal domain decomposition.}
	\label{fig:seismic_wave:subdomain}
\end{figure}

In this part, we present the sequential training scheme via domain decomposition. The reasons for applying domain decomposition are two-fold. Firstly, the scientific problems in geophysics are usually defined in large physical domains, especially for seismic wave propagation. Thus, the computational cost of numerically estimating the solutions of such tasks could be extremely high, which poses a major challenge to PINNs. In addition, the global approximation property of PINNs would lead to a heavy burden on the computational memory of computers/servers. To overcome these computation bottlenecks in large-scale scientific problems, we propose a scalable sequential training strategy via domain decomposition, as shown in Fig.~\ref{fig:seismic_wave:subdomain}. The domain decomposition approach has been employed in prior research works \cite{jagtap2020extended,jagtap2020conservative,kharazmi2021hp} within spatial domains, where interface points are sampled to constrain the continuity of solutions. Nevertheless, different from the aforementioned strategy, we adopt domain decomposition along the time dimension in this paper. This is because we observe more regularity in temporal interfaces compared with spatial domains. To be more specific, we keep the temporal interfaces between two sub-domains identical, which facilitates the generation of the interface points used for ``stitching'' PINNs. Furthermore, a particular PINN is employed to approximate the solution of each sub-domain, and the final global solution is obtained by stitching each local solution from every PINN together. 

As discussed in \cite{jagtap2020conservative}, there exist several benefits of using the domain decomposition scheme, such as better representation capability, improved accuracy of solution due to the local approximation, the flexibility of choosing hyperparameters for each sub-domain, and the convenience for multi-GPU parallelization. However, these benefits are enjoyed at the cost of more coding work. Specifically, an additional loss component $\mathcal{L}_\mathcal{T}$, interface loss that computes the discrepancy of solutions on interfaces from adjacent sub-domains, is included to guarantee the continuity of solutions on the interfaces, e.g., $\mathcal{T}_i$ ($i=1,2,3$) in Fig.~\ref{fig:seismic_wave:subdomain}. Hence, considering $m$ subdomains, the overall loss function can be re-written as
\begin{equation}
\label{eq:loss_func_1}
\mathcal{L}(\mathbf{W},\mathbf{b}) = \eta_0\underbrace{\sum_{i=1}^{m}\mathcal{L}_{e}^i(\mathbf{W}_i,\mathbf{b}_i)}_{\text{equation loss}} + \eta_1\underbrace{\sum_{i=1}^{m}\mathcal{L}_{c}^i(\mathbf{W}_i,\mathbf{b}_i)}_{\text{I/BC loss}}  + \eta_2\underbrace{\sum_{i=1}^{m-1}\mathcal{L}_\mathcal{T}^i(\mathbf{W}_i,\mathbf{W}_{i+1},\mathbf{b}_i,\mathbf{b}_{i+1})}_{\text{interface loss}},
\end{equation}
where $\eta$'s denote the loss weighting coefficients; $\mathcal{L}_{e}^i$ and $\mathcal{L}_{c}^i$ represent the residual loss functions of governing equations and I/BCs, respectively, for the $i$-th sub-domain. The loss function for the $i$-th interface is expressed as
\begin{equation}
\label{eq:loss_func_interface}
\mathcal{L}_\mathcal{T}^i(\mathbf{W}_i,\mathbf{W}_{i+1},\mathbf{b}_i,\mathbf{b}_{i+1}) = \sum_{j=1}^{N_\mathcal{T}}\big\|\mathbf{z}_\mathcal{T}^{i+1}(\mathbf{W}_{i+1},\mathbf{b}_{i+1}; t_j, \mathbf{x}_j) - \mathbf{z}_\mathcal{T}^{i}(\mathbf{W}_{i},\mathbf{b}_{i}; t_j, \mathbf{x}_j)\big\|_2^2,
\end{equation}
Here, $\{t_j, \mathbf{x}_j\}$ denotes the $j$-th collocation point on the interface, $j = 1, 2, ..., N_\mathcal{T}$; $\mathbf{z}_\mathcal{T}^{i}$ and $\mathbf{z}_\mathcal{T}^{i+1}$ represent the physical variables, e.g., $\{\mathbf{u}, \mathbf{v}, \boldsymbol{\sigma}\}$, for the $i$-th and the ($i+1$)-th sub-domains, respectively. Note that $\mathbf{z}$ can be also augmented to account for the time/spatial derivatives of the physical variables to further enhance the continuity condition at the interface.

In genercal, our decomposition strategy on the time domain is easy to implement and retains excellent performance in forward analysis of seismic wave propagation. More details on numerical experiments and results can be found in Section~\ref{s:experiments}.

\subsection{Surrogate modeling for parametric loading}\label{s:surrogate}
Previously, Sun \emph{et al.}~\cite{sun2020surrogate} presented a new direction for surrogate modeling of fluid flows using PINNs. However, the potential of PINNs for surrogate modeling in the field of solid mechanics is still under-explored though the vanilla PINN~\cite{raissi2019physics} shows promises for forward and inverse analysis of various PDE systems. To this end, we propose a new PINN strategy for surrogate modeling of elastic wave propagation. Our goal is to directly predict the corresponding seismic responses under different loading scenarios based on the well-trained PINN model. In particular, we consider different loading locations while keeping the load profie the same. This is commonly seen in seismic wave inversion at the engineering scale (e.g., collecting waveforms given a pre-defined load profile applied at different locations on the Earth surface \cite{trinh2019efficient}). 

\begin{figure}[t!]
	\centering
	\begin{overpic}[width=0.99\textwidth]{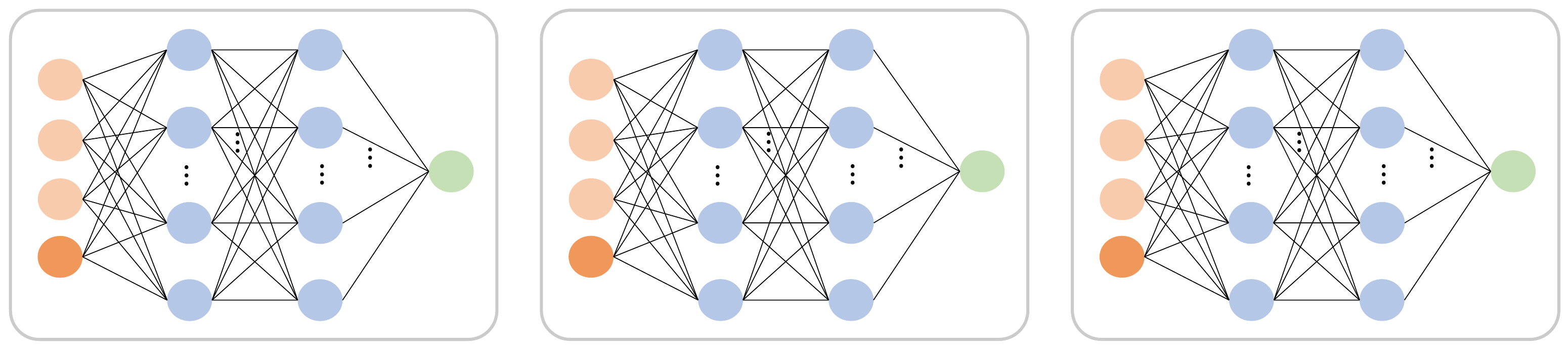}
	
	\put(3.5,16.5){\color{black}{\small $t$}}
	\put(3.3,12.9){\color{black}{\small $x$}}
	\put(3.3,9.3){\color{black}{\small $y$}}
	\put(3.0,5.5){\color{black}{\small $x_c$}}
	\put(28.1,10.8){\color{black}{\small $\boldsymbol{\sigma}$}}

	\put(37.4,16.5){\color{black}{\small $t$}}
	\put(37.2,12.9){\color{black}{\small $x$}}
	\put(37.2,9.3){\color{black}{\small $y$}}
	\put(36.9,5.5){\color{black}{\small $x_c$}}
	\put(62.0,10.8){\color{black}{\small $\mathbf{v}$}}
	
	\put(71.3,16.5){\color{black}{\small $t$}}
	\put(71.1,12.9){\color{black}{\small $x$}}
	\put(71.1,9.3){\color{black}{\small $y$}}
	\put(70.8,5.5){\color{black}{\small $x_c$}}
	\put(95.9,10.8){\color{black}{\small $\mathbf{u}$}}
 	
	\end{overpic}
	\caption{The DNNs part of our proposed surrogate modeling scheme. Here we only consider the spatial variance in $x$-dimension of a 2D domain, which means $\mathbf{x}_c = [x_c,y_c=0]$. In addition, $\{t,x,y\}$ are spatiotemporal coordinates and $\{\boldsymbol{\sigma},\mathbf{v},\mathbf{u}\}$ denote the output variables. There are three different DNNs for learning the corresponding solution variables.}
	\label{fig:surrogate_model}
\end{figure}
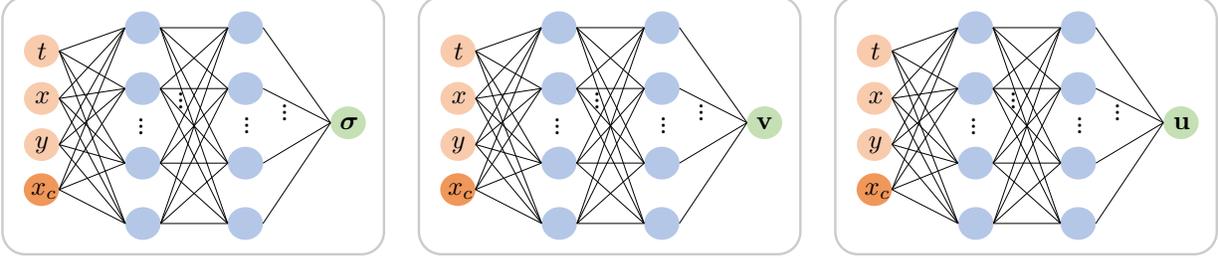

The overview of our designed surrogate modeling scheme is illustrated in Fig.~\ref{fig:surrogate_model}. Let us define a new variable $\mathbf{x}_c$ to represent the spatial coordinate of the loading position. $\mathbf{x}_c$ is also considered as an input variable apart from the spatiotemporal coordinate $\{x,y,t\}$ used in Section~\ref{s:network}. Then, PINNs are utilized to approximate the solutions with respect to the loading scenario in the training stage. Note that $\mathbf{x}_c$ is also directly input to construct the residual loss of the essential BC on the loading area. After training the network, we directly infer the specific seismic responses by changing $\mathbf{x}_c$ to other loading positions of interest. The inspiration of this scheme is that the loading positions at training and extrapolation stages share linear relationships and DNNs are capable of capturing the solution variances caused by the change of spatial information.

Noteworthy, there may be multiple coordinate points for extracting the loading information. For example, for a Gaussian distribution load in 1D domain, it is reasonable to consider three loading points (i.e., $\mathbf{x}_c = [x_{c}^1, x_{c}^2, x_{c}^3]$) as input information due to the complexity and irregularity of such loading information. However, it is also worthwhile to mention that the computational cost increases drastically when inputting more loading points.

\section{Numerical experiments and results}\label{s:experiments}
In this section, a set of numerical examples are implemented to evaluate the capability of our designed PINN architectures for modeling elastic wave propagation in semi-infinite domain. We consider 2D subsurface for all the numerical experiments. In particular, two distinctive material distributions (i.e., Case 1 and Case 2) are designed for solving the elastic wave equations within a truncated domain, and a parametric loading experiment (i.e., Case 3) is displayed for validating the effectiveness of surrogate modeling. Moreover, the ABC described in Section~\ref{s:abc} is adopted in PINNs to avoid wave reflection near the truncated boundaries. We also present the corresponding finite element solutions simulated in enlarged domains to examine the accuracy of PINN results. All the numerical experiments are programmed in TensorFlow~\cite{abadi2016tensorflow} and conducted on an NVIDIA Tesla V100 GPU card (32G) in a standard workstation. 

\subsection{Domain definition}\label{s:domain}
To simplify, we consider the same physical domain and loading scenario for three numerical cases in this paper. The subsurface is excited by the normal traction of the Ricker wavelet type on the surface, which concentrates near the mid-line of the domain (i.e., $x \in [-8,8]$, unit: [m]). The analytical expression of the normal stress applied on the surface is
\begin{equation} 
\label{eq:seismi:normal_stress} 
    \overline{\mathbf{t}}_n(x,0,t)=T_0 \left [ 1-2\pi^2(t-t_s)^2 F_r^2 \right ]\text{e}^{-\pi^2(t-t_s)^2 F_r^2 - (\frac{x}{L})^2} \; \text{[MPa]},
\end{equation}
where $T_0$ is the amplitude of the normal traction, $L$ is the length scale of the normal traction, $t_s$ is the offset of the Ricker wavelet and $F_r$ is the frequency of the excitation. To be more specific, we define $T_0=2.0$ [MPa], $L=1.8$ [m], $t_s=0.1$ [s] and $F_r=15$ [Hz]. Moreover, the material density $\rho$ and Poisson's ratio $\nu$ are set to be 0.002 $\text{[g/mm}^3\text{]}$ and 0.25, respectively. 

Furthermore, the computational domain is truncated into $x\times y\in[-25,25]\times [0,50]$ (unit: [m]) to avoid unnecessary computation. In addition to the surface under normal traction, the remaining three edges are modeled as absorbing boundaries with the ABC as shown in Eq.~\eref{eq:seismi:abc}. Note that we adopt the unit system of [mm], [MPa], and [ms] in the simulation to ensure the output variables (i.e., displacement, velocity, and stress) having similar numerical scales. What's more, the time duration for Case 1 and Case 2 is defined as 400 [ms]. For parametric loading, we reduce it to 200 [ms] since our main goal is to evaluate the performance of our framework in the context of surrogate modeling. It is noteworthy that the material distribution for each case is different and described as follows. 

\subsection{Evaluation metric}\label{s:metric}
We define an accumulative root-mean-square error (a-RMSE) to measure the error propagation between the learned waveforms and the ground truth, expressed as 
\begin{equation} 
    \label{eq:accum_rmse} 
    \begin{aligned}
    \text{a-RMSE}(t_i) = \sqrt{\text{MSE} \left ( \widehat{\boldsymbol{\mathcal{U}}}^{(1:i)}-\boldsymbol{\mathcal{U}}_{\text{ref}}^{(1:i)}\right )}
    \end{aligned}
\end{equation}
where $\widehat{\boldsymbol{\mathcal{U}}}^{(1:i)}$ and $\boldsymbol{\mathcal{U}}_{\text{ref}}^{(1:i)}$ denote the predicted and the reference solutions, respectively, from the $1^\text{st}$ to the $i^\text{th}$ step. It describes the misfit of all snapshots before the time instance $t_i$.

\begin{figure}[t!]
	\centering
	\includegraphics[width=0.7\textwidth]{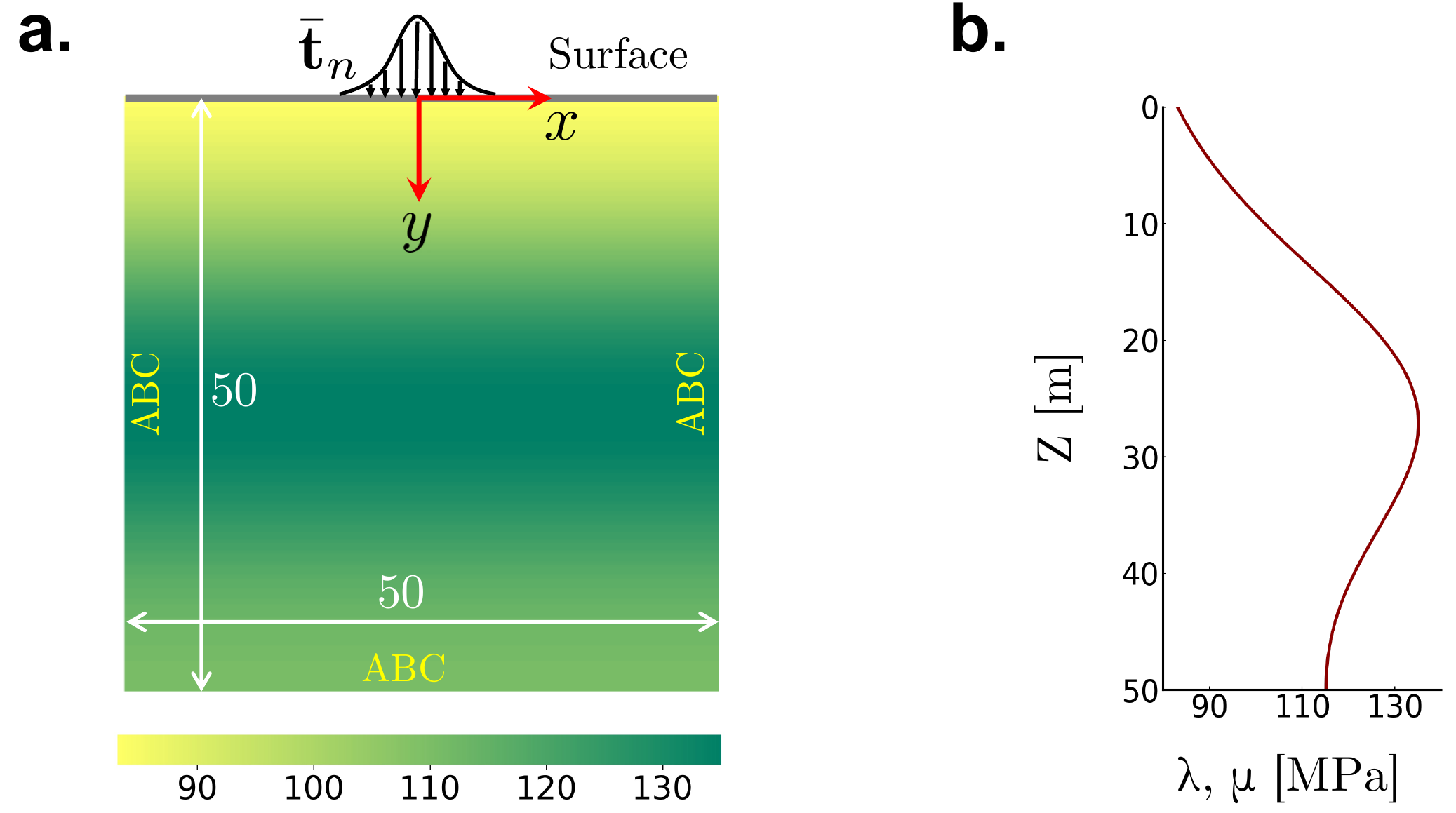}
	\caption{The computational domain with a mixed material distribution of linear and Gaussian functions. $\mathbf{a.}$ The spatial domain is defined with $x\times y\in[-25,25]\times[0,50]$ (unit: [m]). ABC denotes the absorbing boundary condition. $\mathbf{b.}$ The material distribution is presented along the depth $y$.}
	\label{fig:linear_gaussian_dist}
\end{figure}

\begin{figure}[t!]
\centering
\includegraphics[width=0.99\textwidth]{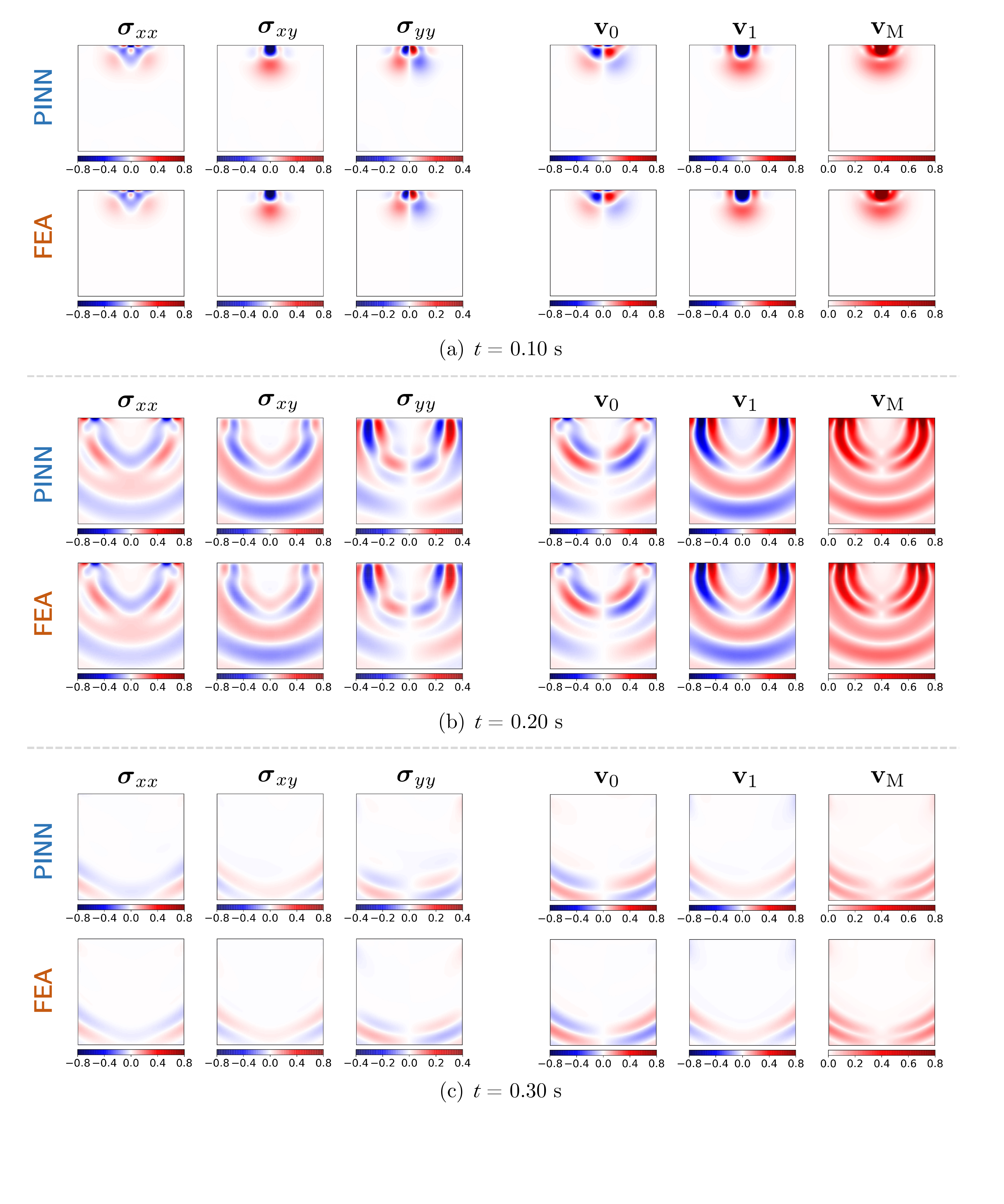}
\caption{The results of stress and velocity distributions obtained from our proposed PINN and FEA in Case 1. We select three representative snapshots (i.e., $t=0.10, 0.20, 0.30$) for comparison. $\mathbf{v}_0$, $\mathbf{v}_1$, and $\mathbf{v}_\text{M}$ denote the horizontal velocity, the vertical velocity, and the magnitude respectively.}
\label{fig:case1_snapshots}
\end{figure}

\begin{figure}[t!]
\centering
\includegraphics[width=0.95\textwidth]{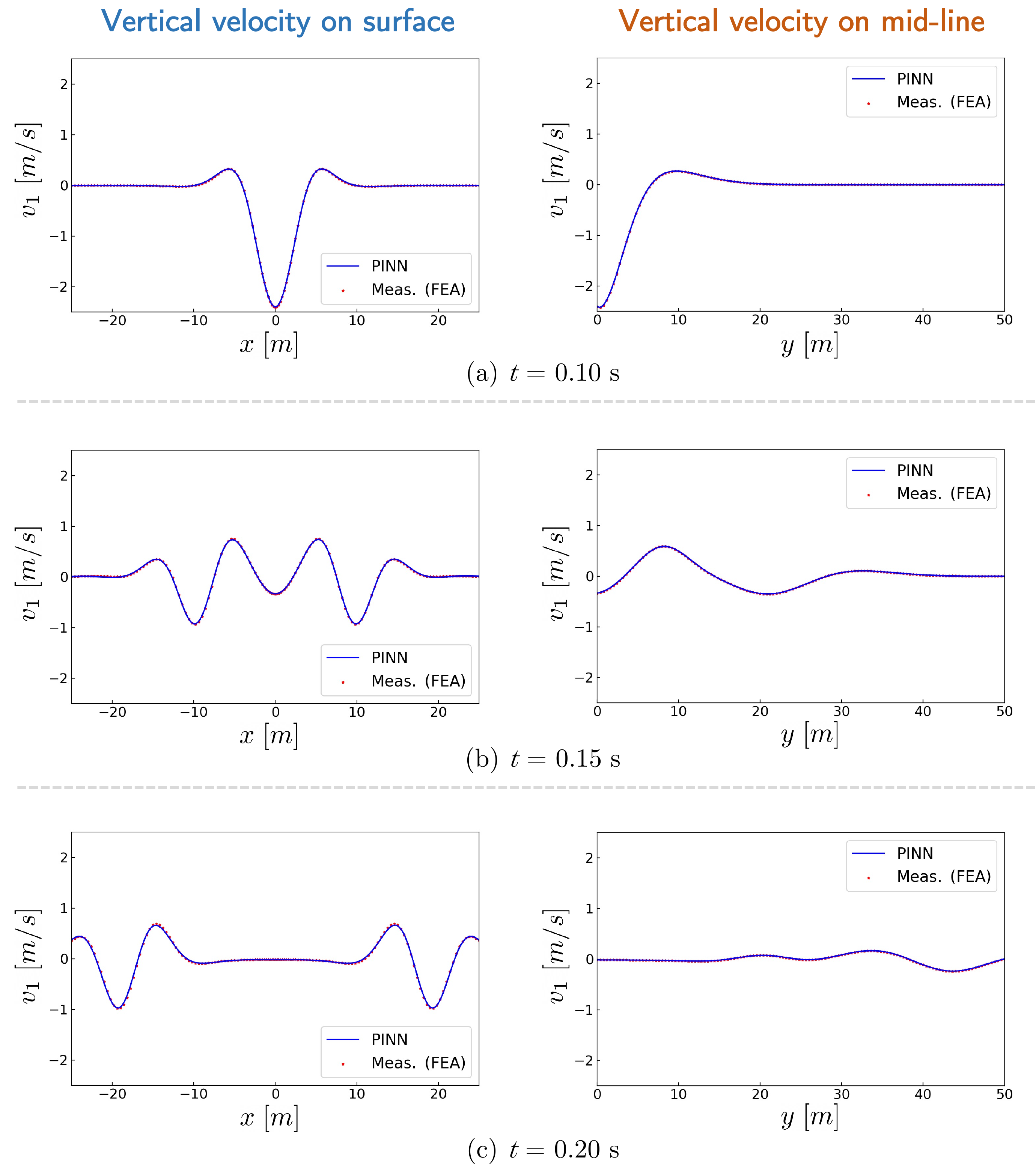}
\caption{Vertical velocity distributions (i.e., $\mathbf{v}_1$) at various time steps on the surface ($y=0$, left) and mid-line ($x=0$, right) in Case 1. The synthetic measurement is obtained from FEA on an enlarged domain ($x\times y\in[-75,75]\times [0,100]$, unit: [m]).}
\label{fig:case1_vel}
\end{figure}

\begin{figure}[t!]
\centering
\includegraphics[width=0.99\textwidth]{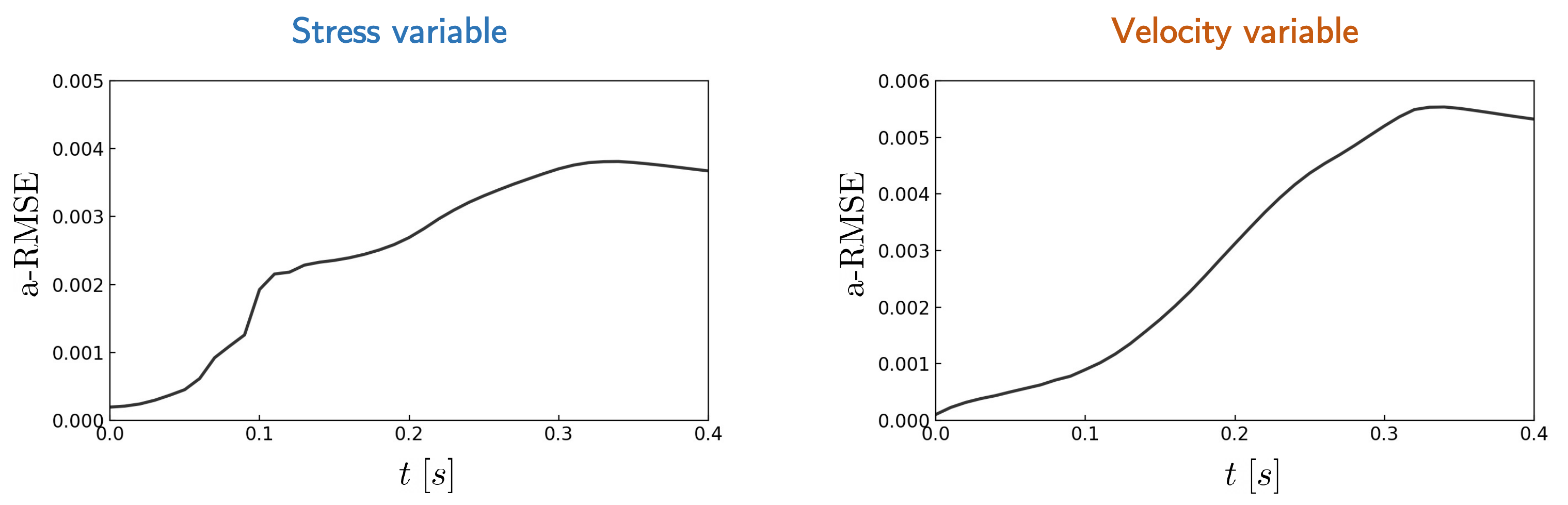}
\caption{Error propagation of our proposed PINN framework for stress and velocity variables in Case 1.}
\label{fig:case1_err_prop}
\end{figure}

\subsection{Solving wave propagation equations}\label{s:solving_wave_eqn}
\subsubsection{Case 1 -- Linear and Gaussian material distribution}\label{s:linear_gaussian}
In the first example, we consider a 2D subsurface with the material property following a combination of linear and Gaussian functions, where the Lam\'e constants of the media are given by
\begin{equation} 
\label{gauss_distr} 
\lambda(x,y) = \mu(x,y)=80+0.64y+38.4 \mathrm{e}^{-\frac{(y-25)^2}{250}} \; \text{[MPa]}.
\end{equation}
The graphic illustration of the physical domain and the material distribution is exhibited in Fig.~\ref{fig:linear_gaussian_dist}. Note that this material distribution is relatively simple and smooth. Therefore, domain decomposition is not applied here since there is no harsh requirement for network size and collocation points. In specific, we employ two deep networks of $[16+3\times80]$\footnote{It denotes the hidden layers with 16, 80, 80 and 80 neurons respectively.} for learning the displacement $\mathbf{u}$ and the velocity $\mathbf{v}$. Moreover, another independent network of $[24+3\times120]$ is designed for approximating the stress variable $\bm{\sigma}$. We sample $218,084$ collocation points within the physical domain to enforce the governing equations, among which $50,000$ points are near the wave source. In addition, $10,000$, $58,431$, and $12,000$ points are sampled on IC, each absorbing boundary and the surface boundary, respectively. The sampling method for collocation points is slightly different from the Latin hypercube sampling (LHS) strategy~\cite{mckay2000comparison}, which is typically used in PINN applications. The related details are elaborated in \ref{s:sampling}.

Furthermore, we apply Adam~\cite{kingma2014adam} and L-BFGS-B~\cite{zhu1997algorithm} optimizers to sequentially train the networks. In Case 1, $2,000$ Adam and $20,000$ L-BFGS-B iterations are used. Note that we consider the sine activation function~\cite{sitzmann2020implicit} in the networks. The selection of hyper-parameters is described in \ref{s:hyper-para}. What's more, to generate the reference solution, we also perform the numerical simulation of this problem by FEA using an enlarged computational domain (i.e., $x\times y\in[-75,75]\times [0,100]$, unit: [m]), where the waves propagate out the truncated domain naturally. 

The comparative results of snapshots are shown in Fig.~\ref{fig:case1_snapshots}. It is obvious that the solution of PINN matches the ground truth very well. The excellent performance is further validated by comparing the vertical velocity distributions on the surface (i.e., $y=0$ [m]) and the mid-line (i.e., $x=0$ [m]). The zoom-in results also exhibit a remarkable agreement between the prediction of PINN and the reference solution. In addition, we showcase the error propagation curves of stress $\boldsymbol{\sigma}$ and velocity $\mathbf{v}$ in Fig.~\ref{fig:case1_err_prop}. The errors increase mildly along with time marching and remain at a small-error level (i.e., $0.005$). For Case 1 with a relatively simple material distribution (i.e., linear and Gaussian function), all of the numerical results prove the excellent solution accuracy of our proposed PINN architecture.

\subsubsection{Case 2 -- Two-layer material distribution}\label{s:2layer_material}

\begin{figure}[t!]
	\centering
	\includegraphics[width=0.7\textwidth]{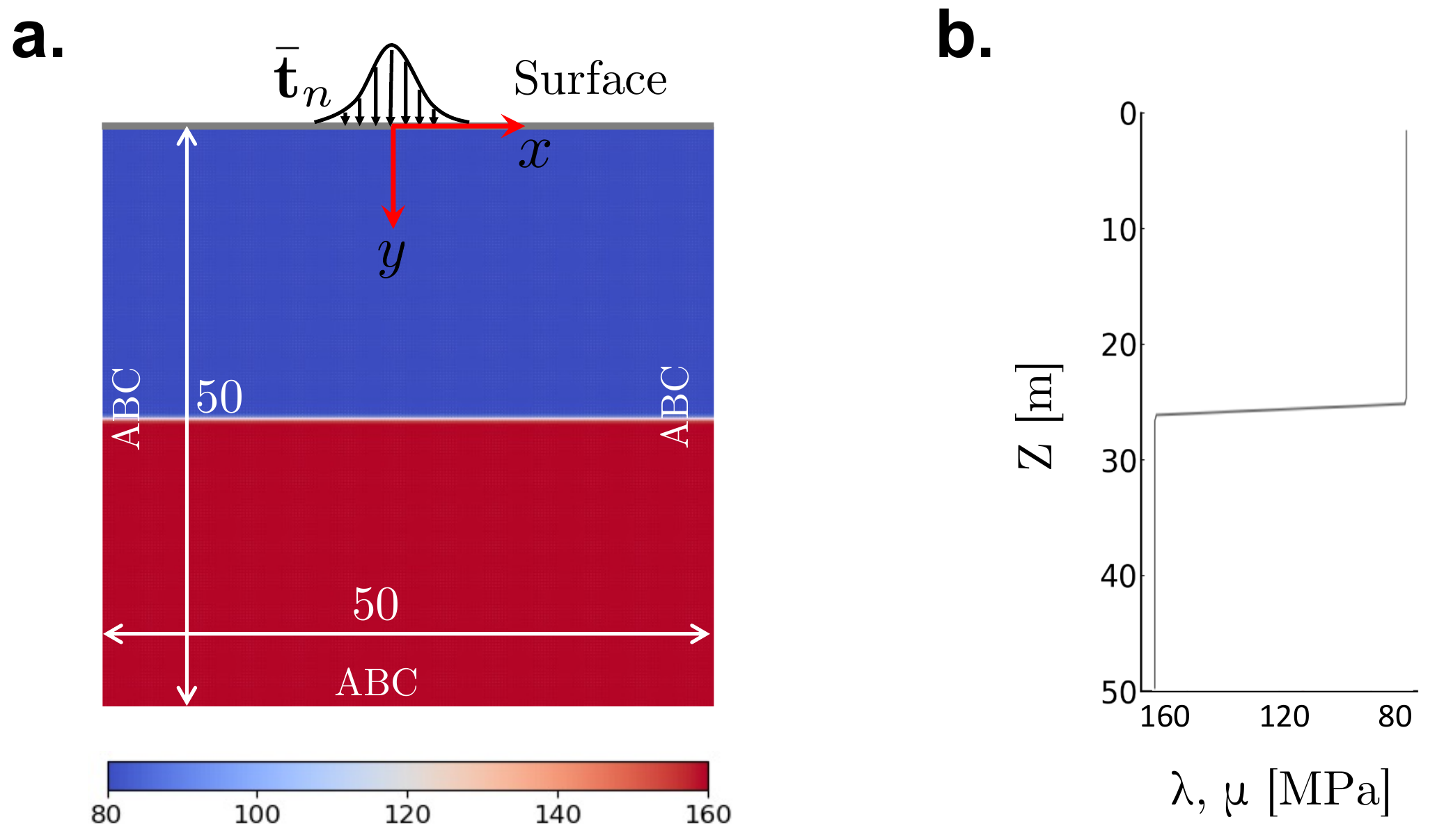}
	\caption{The illustrative diagram of the 2D subsurface with two layers of media. $\mathbf{a.}$ The spatial domain is defined with $x\times y\in[-25,25]\times[0,50]$ (unit: [m]). ABC denotes the absorbing boundary condition. $\mathbf{b.}$ The material distribution is presented along the depth $y$.}
	\label{fig:seismic_wave:diagram_2layers}
\end{figure}

\begin{figure}[t!]
\centering
\includegraphics[width=0.99\textwidth]{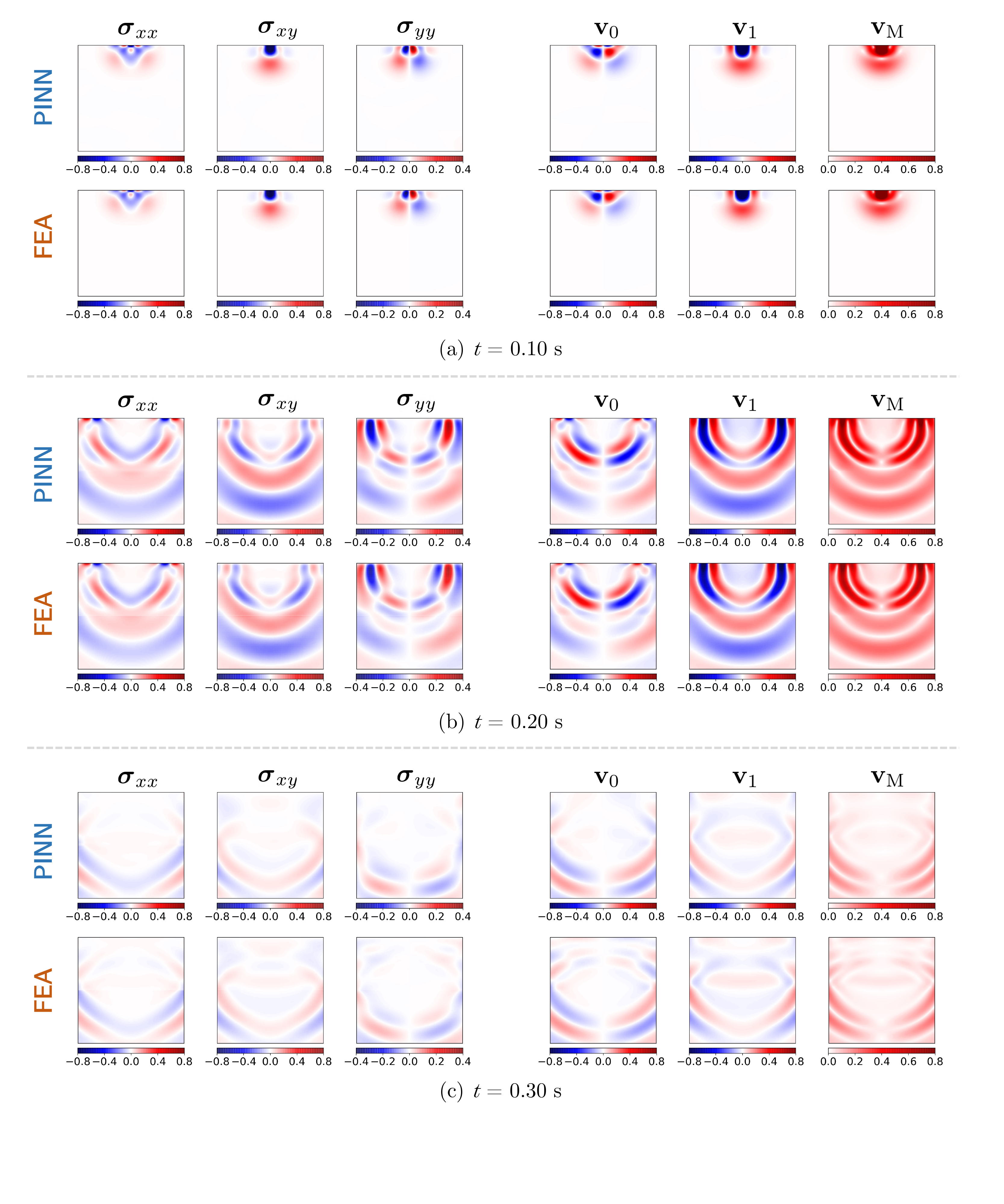}
\caption{The results of stress and velocity distributions obtained from our proposed PINN and FEA in Case 2. We select three representative snapshots (i.e., $t=0.10, 0.20, 0.30$ [s]) for comparison. $\mathbf{v}_0$, $\mathbf{v}_1$, and $\mathbf{v}_\text{M}$ denote the horizontal velocity, the vertical velocity, and the magnitude respectively.}
\label{fig:case2_snapshots}
\end{figure}

\begin{figure}[t!]
\centering
\includegraphics[width=0.95\textwidth]{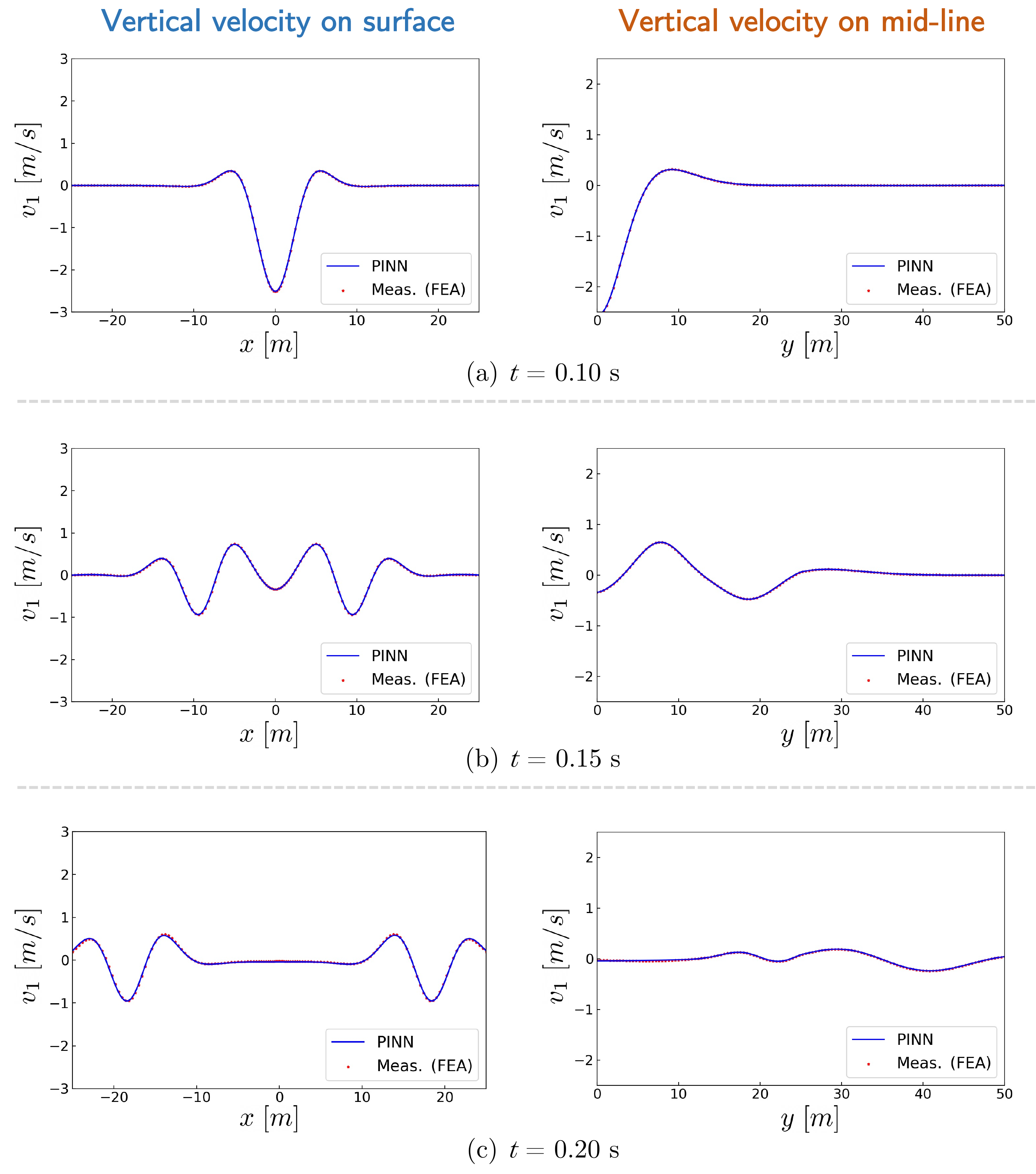}
\caption{Vertical velocity distributions (i.e., $\mathbf{v}_1$) at various time steps on the surface ($y=0$, left) and mid-line ($x=0$, right) in Case 2. To obtain the PINN prediction, the truncated domain is decomposed into two parts along the time dimension. Each part is modeled with a separate PINN.}
\label{fig:case2_vel}
\end{figure}

\begin{figure}[t!]
\centering
\includegraphics[width=0.99\textwidth]{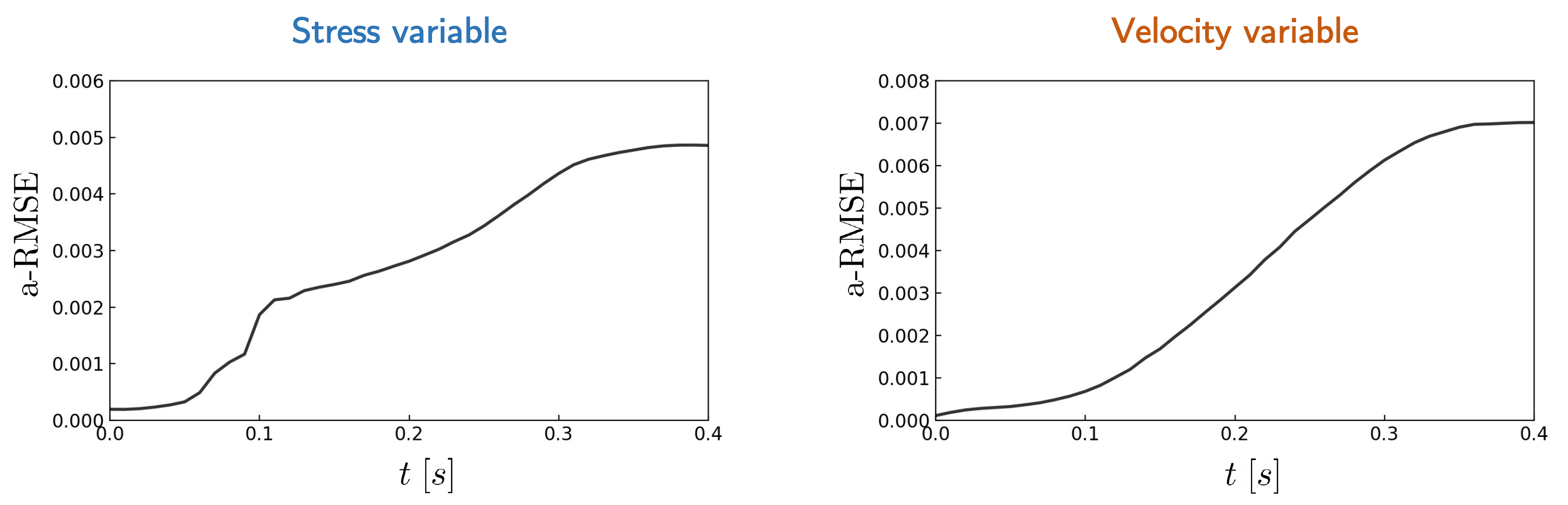}
\caption{Error propagation of our proposed PINN framework for stress and velocity variables in Case 2.}
\label{fig:case2_err_prop}
\end{figure}

The second numerical case is a two-layer subsurface as shown in Fig.~\ref{fig:seismic_wave:diagram_2layers}. More precisely, we prescribe the material property of the subsurface as
\begin{equation} 
\label{eq:seismi:mat_dist} 
    \lambda(x,y) = \mu(x,y) =80+\frac{80}{1+\text{e}^{-10(y-25)}},
\end{equation}
which is under the assumption of elasticity. In this case, due to the complexity of the material property, we consider using more collocation points than those in Case 1. Hence, in spite of being truncated, the computational domain is still very large and cannot fit into the memory (i.e., 32GB) of one single Tesla V100 GPU. Therefore, we decompose the spatiotemporal domain into two parts along the time dimension (i.e., $t\in[0,0.2]\cup [0.2,0.4]$, unit: [s]). The solution of each subdomain is approximated by a separate PINN as described in Sub-section \ref{s:domain_decomp}.

Herein, we also design two networks of $[16+3\times80]$ for the variables of displacement $\mathbf{u}$ and the velocity $\mathbf{v}$, as well as a network of $[24+3\times120]$ for the stress variable $\bm{\sigma}$. Besides, $200,188$ collocation points are sampled within the domain to evaluate the residual of the governing equations. $4,000$ points are sampled on each absorbing boundary while $8,000$ points are generated on the surface. In addition, $6,135$ interface points are used to guarantee the continuity of the solution. Each PINN is trained with $2,000$ epochs of Adam and $60,000$ iterations of L-BFGS-B. Note that the configuration introduced in this part is specified for each sub-domain.

Fig.~\ref{fig:case2_snapshots} presents the snapshots of the velocity and stress fields predicted by PINN and FEA at different time steps. It can be seen that the overall prediction by PINN agrees with the reference solution (i.e., FEA) very well. It is interesting to point out that for $t=0.30$ [s], the reference solution of FEA is characterized by more details on the reflected wave compared with the learned solution by PINN. This is because the reflected wave has a much smaller scale than the incident wave, which poses a major challenge to the optimization of the loss function. This observation implies the limitation of the point-wise global approximation property of PINN. When scientific problems feature fine-scale patterns, the PINN that leverages fully-connected NNs as approximator might be unable to capture some delicate phenomena. The comparisons of the velocity distribution on the surface (i.e., $y=0$ [m]) and mid-line (i.e., $x=0$ [m]) are shown in Fig.~\ref{fig:case2_vel} to quantitatively examine the accuracy of the proposed PINN. We observe that the prediction of PINN matches perfectly with the reference solution, which demonstrates its excellent accuracy. Moreover, the error propagation is displayed in Fig.~\ref{fig:case2_err_prop}. The error evolutions are relatively larger than those of Case 1 due to a more complex material distribution, but the solution performance is still satisfactory thanks to the domain decomposition strategy.

\begin{figure}[t!]
\centering
\includegraphics[width=0.35\textwidth]{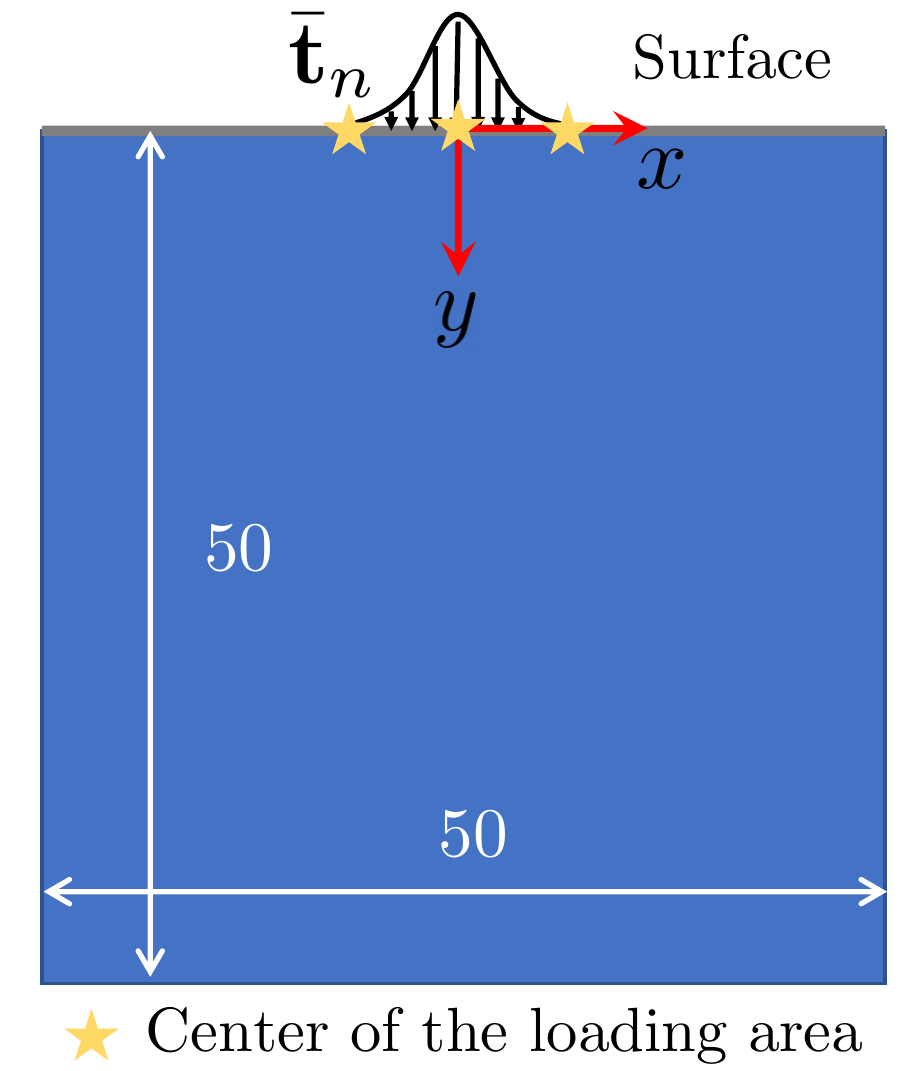}
\caption{The illustrative diagram of the 2D subsurface with homogeneous material for parametric loading. The yellow stars denote the centers of the loading areas. The truncated boundaries are regarded as free boundaries due to the setting of homogenous material. The absorbing boundary condition is not required here.}
\label{fig:para_loading_domain}
\end{figure}

\begin{figure}[htbp]
\centering
\includegraphics[width=0.84\textwidth]{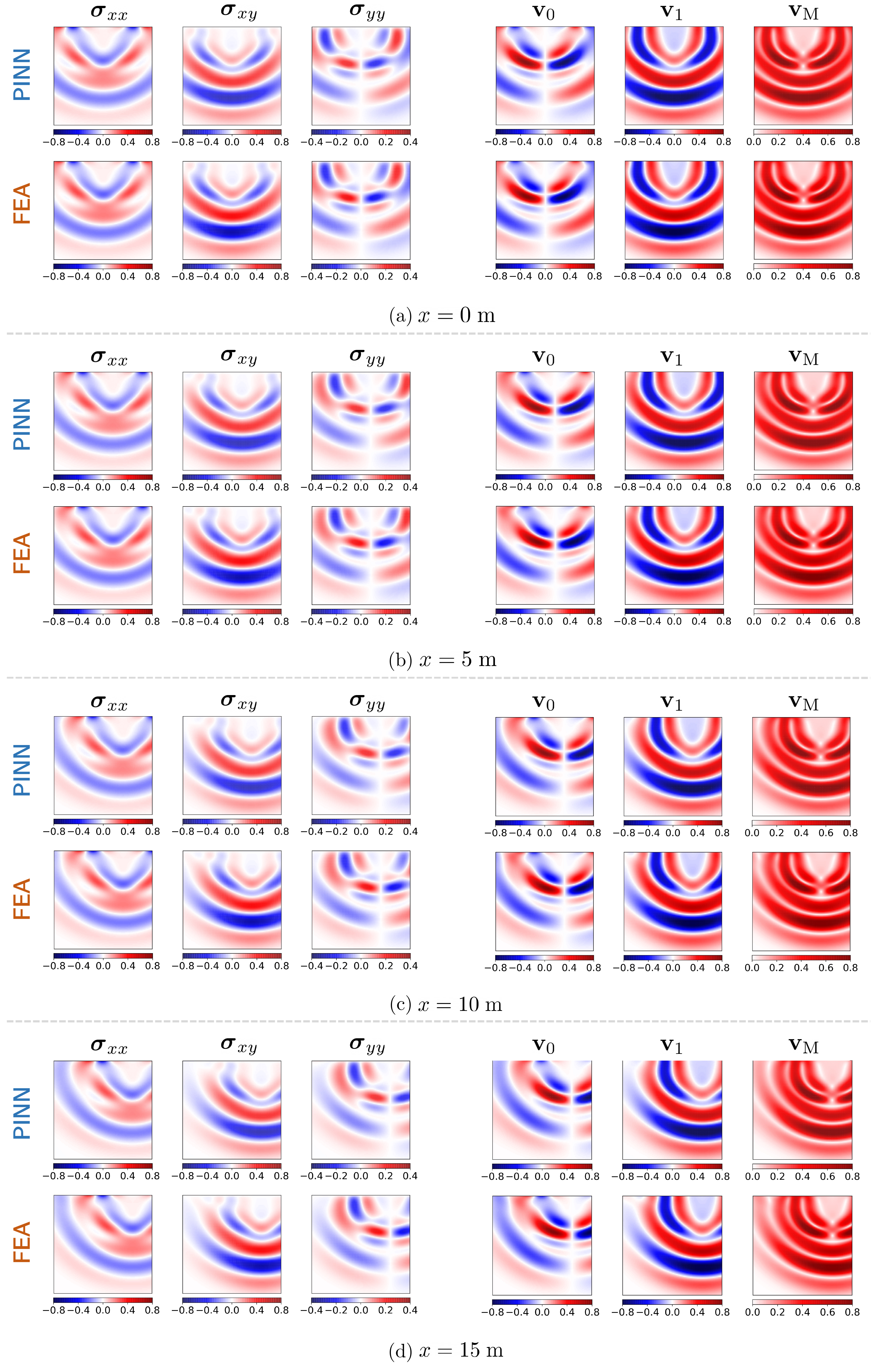}
\caption{The snapshots of stress and velocity from our proposed PINN and FEA in Case 3. We present two interpolation ($x=0,5$ [m]) and two extrapolation ($x=10,15$ [m]) results. The time step is select at $t=0.2$ [s].}
\label{fig:case3_snapshots}
\end{figure}

\begin{figure}[htbp]
\centering
\includegraphics[width=0.99\textwidth]{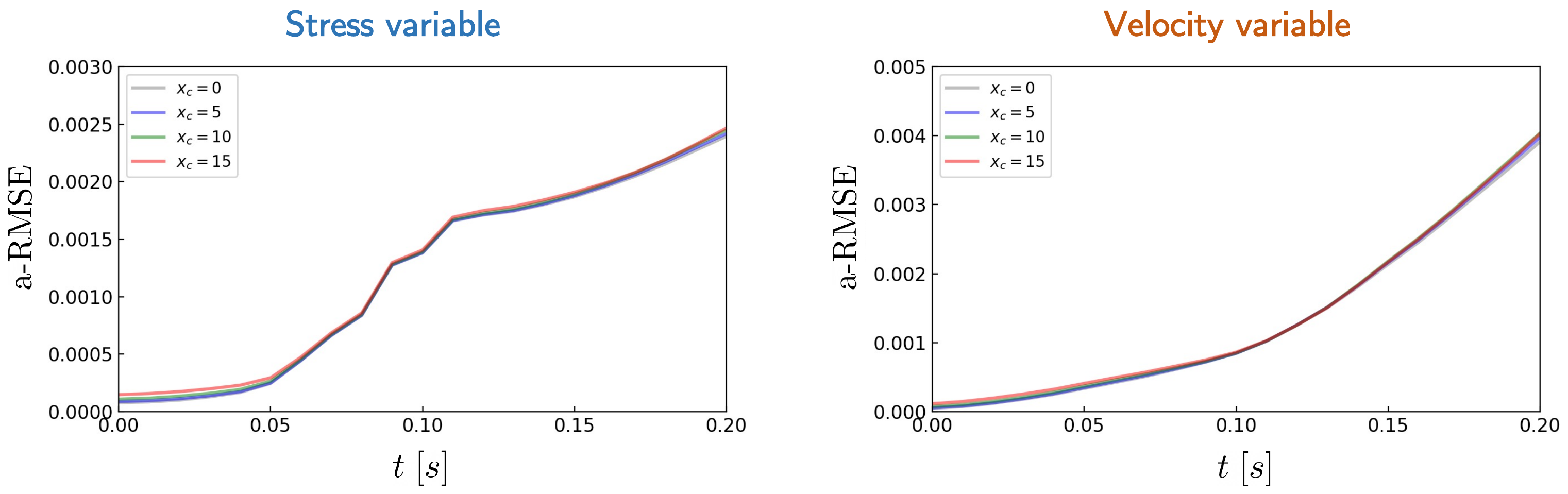}
\caption{Error propagation of our proposed PINN framework for stress and velocity variables in Case 3. Four loading scenarios are selected in the diagram, i.e., $x_c=0,5,10,15$ [m].}
\label{fig:case3_err_prop}
\end{figure}

\subsection{Case 3 -- Parametric loading}


For the parametric loading case, we consider a homogeneous material in the physical domain, where the Young's modulus $E$ is defined as $200$ [MPa]. The spatial domain size is also set as $x\times y\in[-25,25]\times [0,50]$ (unit: [m]). 
The time duration for this case is 0.2 [s]. The domain decomposition is not applied here due to the relatively small domain size. Moreover, the loading follows the formulation of Eq.~\eref{eq:seismi:normal_stress} with $T_0=-2.0$ [MPa], $L=5.0$ [m]. However, different from solving seismic wave propagation, we incorporate three excitation positions in $\mathbf{x}_c$ (e.g., $[-5,0,5]$, unit: [m]) in addition to the spatiotemporal coordinate information (i.e., $\{x,y,t\}$), which works as loading parameters for surrogate modeling. Each location is defined as the center of loading area.

Three separate fully-connected NNs of $5\times100$ are used to approximate the three variables $\{\mathbf{u}, \mathbf{v}, \boldsymbol{\sigma}\}$. We sample $152,043$ collocation points over the spatiotemporal domain and additional $50,000$ points near the wave source to evaluate the residual of the governing equations. Moreover, $10,500$ and $88,236$ points are sampled for IC and the surface boundary condition, respectively. We train the network with $2,000$ Adam iterations and $80,000$ L-BFGS-B epochs until convergence. In specific, we define three loading positions of $\mathbf{x}_c=\{-5,0,5\}$ (unit: [m]) to train the network. Based on the trained model, we directly infer/extrapolate the dynamics under other loading positions (e.g., $\mathbf{x}_c=10,15$ [m]).

The results of parametric loading are shown in Figs.~\ref{fig:case3_snapshots} and \ref{fig:case3_err_prop}. We show representative snapshots (i.e., $t=0.2$ [s]) of two training ($x_c=0,5$ [m]) and two extrapolation scenarios ($x_c=10,15$ [m]). As shown in Fig.~\ref{fig:case3_snapshots}, both the training and extrapolation results from our proposed PINN match the corresponding ground truth well. It is worthwhile to mention that the extrapolation of $x_c=15$ [m] maintains excellent solution accuracy, which proves the capability of our method for surrogate modeling. In addition, the error propagation in Fig.~\ref{fig:case3_err_prop} further validates the great performance in the context of solving the wave equations under parametric loading. In particular, the error propagations of extrapolation present similar evolving patterns as those of the training scenarios. The errors of the stress variable increase softly, while the errors of the velocity variable show relatively large increments but still acceptable.

\section{Discussions}\label{s:discussion}
We herein discuss the current limitations of the proposed PINN framework and the outlook of our future work. Generally, the main potential concerns come from (1) the soft imposition scheme of enforcing I/BCs and (2) the scalability issue due to the setting of fully-connected NNs. More precisely, the typical PINNs consider the physical principles as loss regularizer terms in the optimization. However, it is challenging for such an implementation to guarantee the learned dynamics rigorously following the underlying physical laws. This observation is empirically validated by a recent work \cite{krishnapriyan2021characterizing} which analyzes the failure modes of PINNs. Therefore, instead of using a soft enforcement strategy, we would like to explore the possibility of encoding the known physics (e.g., the PDE structure) into the networks~\cite{rao2022discovering}. In addition, PINNs employ fully-connected NNs for approximating the solution of PDEs, which can be essentially regarded as a continuous learning scheme. They are excellent in capturing global patterns but show inferior performance in learning local details compared with discrete learning methods (e.g., convolution-based NNs)~\cite{ren2022phycrnet}. Moreover, applying discrete learning has potential to mitigate the computation burden of learning seismic wave propagation in large domains. In the future, we will put more research efforts into building discrete learning frameworks (e.g., convolutional and graph NNs) for forward analysis of spatiotemporal dynamics. In addition, our future work will also be placed on extending the proposed model to solve full wave inversion problems.

\section{Conclusions}\label{s:conclusion}
In this paper, we propose a new PINN architecture for forward modeling of seismic wave propagation in semi-infinite domain. It is capable of both solving the elastic wave equations and parametric surrogate modeling within truncated domains. There are three characteristics highlighted: (1) the introduction of the ABC into the network as a soft regularizer to eliminate boundary effect and avoid expensive computation in semi-infinite domains; (2) a new sequential training scheme via temporal domain decomposition to improve scalability and solution accuracy; (3) a parametric surrogate modeling scheme to predict the seismic responses under different loading scenarios. Note that the entire network is trained without any labeled data. Furthermore, we evaluate the performance of our proposed PINN architectures on various numerical cases with different material distributions. The results demosntrate the effectiveness of our approach in the context of solution accuracy and extrapolation capability.

\section*{Acknowledgement}
P. Ren would like to gratefully acknowledge the Vilas Mujumdar Fellowship at Northeastern University. Y. Liu would like to thank the support from the Fundamental Research Funds for the Central Universities. In addition, P. Ren also would like to sincerely acknowledge the constructive discussion with Dr. Jin-Yun Xie.

\section*{Data Availability}
All the datasets and source codes to reproduce the results in this study are available on GitHub at \url{https://github.com/paulpuren/seismic_modeling} upon final publication.

\appendix
\section{The sampling of collocation points}\label{s:sampling}
In this paper, we integrate the traditional LHS method and Gaussian Quadrature~\cite{hughes2012finite} in FEA for sampling collocation points. The reason for such a hybrid strategy is that we observe more uniformly distributed collocation points based on Gaussian Quadrature compared with the LHS method. Therefore, we discretize the entire spatiotemporal domain with quadrilateral and then sample Gaussian points. In addition, we consider using the LHS approach as a supplement for adding more collocation points near the source wave.

\section{The selection of hyper-parameters}\label{s:hyper-para}
The general principle of selecting hyper-parameters (i.e., the weighting coefficients for loss terms) is to ensure that each weighted loss term shares a similar numerical scale. In our implementation, we first initialize the entire network with Xavier's method~\cite{glorot2010understanding} and then train the specific PINN with one iteration in order to see the magnitudes of loss terms. Next, the weighting coefficients are defined by making the products of the coefficients and the corresponding loss terms close to ones. The hyper-parameters used in this study are presented in Table~\ref{tab:hyper-para}. We extend the loss function in Eq.~\eref{eq:loss_func} to a more specific formulation, which is defined as 
\begin{equation}
    \label{eq:loss_coeff}
    \mathcal{L} = \eta_0 \mathcal{L}_e + \eta_1 \mathcal{L}_{ic} + \eta_2 \mathcal{L}_s + \eta_3 \mathcal{L}_{v} + \eta_4 \mathcal{L}_{nb} + \eta_5 \mathcal{L}_{src} + \eta_6 \mathcal{L}_{abc},
\end{equation}
where $\{\mathcal{L}_{ic}, \mathcal{L}_s, \mathcal{L}_{v}, \mathcal{L}_{nb}, \mathcal{L}_{src}, \mathcal{L}_{abc}\}$ denote the loss terms for IC, displacement-stress equation, velocity-displacement equation, natural boundary condition, wave source condition and absorbing boundary condition, respectively. $\{\eta_1,\eta_2,\eta_3,\eta_4,\eta_5,\eta_6\}$ are their corresponding hyper-parameters. Note that absorbing boundary condition is not required for homogenous material distribution~\cite{rao2021physics} in Case 3 (i.e., parametric loading experiment).

\begin{table}[t!]
\centering
\caption{The details of hyper-parameters for three numerical cases.``N/A'' represents being inapplicable to the specific index.}
{\small
\begin{tabular}{cccccccc}
\toprule
 & $\eta_0$ & $\eta_1$ & $\eta_2$ & $\eta_3$ & $\eta_4$ & $\eta_5$ & $\eta_6$ \\
\midrule
Case 1 & 10000  & 1 & 10000  & 5 & 10  & 10 & 10000000\\
\midrule
Case 2 & 10000  & 1 & 10000  & 5 & 10  & 10 & 10000000\\
\midrule
Case 3 & 4000000  & 200 & 1000000  & 200 & 400  & 400 & N/A\\
\bottomrule
\end{tabular}
}
\label{tab:hyper-para}
\end{table}

\bibliographystyle{elsarticle-num}
\bibliography{refs.bib}

\end{document}